\documentclass[pra,superscriptaddress,twocolumn]{revtex4}
\usepackage{graphicx}
\usepackage{bm}
\usepackage{amsmath}
\usepackage{amssymb}
\usepackage{exscale}
\usepackage{wasysym}
\usepackage{ulem}
\usepackage{cancel}
\bibstyle{apsrev.bib}

\usepackage{xcolor}

\begin{document}

\title{Microscopic theory for the pair correlation function of liquidlike colloidal suspensions under shear flow}

\author{Luca Banetta}

\thanks{These authors contributed equally to this work}

\affiliation{Department of Applied Science and Technology, Politecnico di Torino, Corso Duca degli Abruzzi 24, 10129, Turin, Italy}

\author{Francesco Leone}

\thanks{These authors contributed equally to this work}

\affiliation{Department of Physics ``A. Pontremoli", University of Milan, via Celoria 16, 20133 Milan, Italy}

\author{Carmine Anzivino}

\thanks{These authors contributed equally to this work}

\affiliation{Department of Physics ``A. Pontremoli", University of Milan, via Celoria 16, 20133 Milan, Italy}

\author{Michael S. Murillo}

\affiliation{Department of Computational Mathematics, Science and Engineering, Michigan State University, East Lansing, Michigan 48824, USA}

\author{Alessio Zaccone}

\email[Electronic mail: \ ]{alessio.zaccone@unimi.it}

\affiliation{Department of Physics ``A. Pontremoli", University of Milan, via Celoria 16, 20133 Milan, Italy}

\date{\today}

\begin{abstract}
We present a theoretical framework to investigate the microscopic structure of concentrated hard-sphere colloidal suspensions under strong shear flows by fully taking into account the boundary-layer structure of convective diffusion. We solve the pair Smoluchowski equation with shear separately in the compressing and extensional sectors of the solid angle, by means of matched asymptotics. A proper, albeit approximate, treatment of the hydrodynamic interactions in the different sectors allows us to construct a potential of mean force containing the effect of the flow field on pair correlations. We insert the obtained pair potential in the Percus-Yevick relation and use the latter as a closure to solve the Ornstein-Zernike integral equation. For a wide range of either the packing fraction $\eta$ and the P\'eclet ($\textrm{Pe}$) number, we compute the pair correlation function and extract scaling laws for its value at contact. For all the considered value of $\textrm{Pe},$ we observe a very good agreement between theoretical findings and numerical results from literature, up to rather large values of $\eta.$ The theory predicts a consistent enhancement of the structure factor $ S(k)$ at $k \to 0,$ upon increasing the $\textrm{Pe}$ number. We argue this behaviour may signal the onset of a phase  transition from the isotropic phase to a non-uniform one, induced by the external shear flow.
\end{abstract}

\maketitle

\section{Introduction}

A long-standing problem in soft condensed matter physics is to determine the microscopic structure of a colloidal suspension as a function of the control parameters, when the interaction potential among the particles in the suspension is known. A possible description for the microscopic structure is given by the so-called \textit{pair correlation function}, $g (\mathbf{r}_{1}, \mathbf{r}_{2}).$ If $N$ is the number of colloidal particles dispersed in the suspension, $g (\mathbf{r}_{1}, \mathbf{r}_{2})$ represents the probability of finding a first particle in a volume $d \mathbf{r}$ centered at $\mathbf{r}_{1},$ and a second particle in a volume $d \mathbf{r}$ centered at $\mathbf{r}_{2},$ irrespective of the position of the remaining $N-2$ particles \citep{Hansen_McDonald_BOOK}. Efficient methods to compute the pair correlation function of a colloidal suspension at equilibrium are either simulations \citep{Allen_Tildesley} and integral equation theories \citep{caccamoIET}. More challenging, instead, is to compute $g (\mathbf{r}_{1}, \mathbf{r}_{2})$ in a system subjected to an external shear flow, a problem which has many relevant applications in rheology \citep{PhysRevLett.101.138301,PhysRevLett.89.248304} and the preparation of nanomaterials \citep{Preziosi,doi:10.1021/la902800x}. In the case of sheared colloidal suspensions, the spatial arrangement of the colloidal particles results from an intricate interplay of interparticle interactions, Brownian motion, shear-induced flow field contributions and hydrodynamic interactions \citep{Vermant_2005}. A so-called P\'eclet number ($\textrm{Pe}$) is typically introduced to describe the relative importance of shear-induced to Brownian effects. For spherical particles of diameter $\sigma,$ the $\textrm{Pe}$ number is defined as 
\begin{equation} \label{definition_Peclet}
\textrm{Pe} = \frac{6 \pi \eta_{0} (\sigma / 2)^{3} \dot{\gamma}}{k_{B} T},
\end{equation} 
where $\eta_{0}$ is the viscosity of the hosting fluid, $\dot{\gamma}$ is the shear rate, while $k_{B}$ and $T$ are the Boltzmann constant and the absolute temperature, respectively. In suspensions with $\textrm{Pe} \gg 1$ ($\textrm{Pe} \ll 1$), the flow field (Brownian motion) is the dominant contribution.

The pair correlation function of a colloidal suspension under shear flow can be obtained by solving the so-called pair Smoluchowski equation with shear \citep{Dhont_BOOK}. Several attempts to solve the latter equation have been proposed in the past decades, in the particular case of hard-sphere colloidal suspensions under strong shear flow, i. e. for $\textrm{Pe} \gg 1.$ Analytical approaches include the exact solution found by Batchelor and Green in the $\textrm{Pe} \to \infty$  limit \citep{batchelor_green_1972}, and the work of Brady and Morris which featured the presence of a \textit{boundary layer} of thickness $\mathcal{O} (\textrm{Pe}^{-1})$  \citep{brady_morris_1997}. For weak shear flows, i. e. $\textrm{Pe} \ll 1,$ it is easier to approach the Smoluchowski equation in the Fourier rather than in the real space \citep{dhont_1989,Szamel,Schwarzl,Ronis}. An account of the shear-induced distortion of the structure factor in colloidal suspensions can be obtained in this case, a phenomenon which has been widely investigated also experimentally \citep{Kruif,Clark,Ackerson}.

A new analytical scheme based on intermediate asymptotics has been recently introduced to solve the pair Smoluchowski equation with shear, separately in the \textit{compressing} and the \textit{extensional} sectors of the solid angle \citep{Banetta_Zaccone_PRE}. While in the compressing sectors the particles are pushed towards each other by the shear flow, in the extensional sectors the particles are pulled away from one another by the shear flow. 
The method introduced in Ref. \citep{Banetta_Zaccone_PRE} can be applied to systems displaying different types of inter-particle interactions, and has been succesfully employed for suspensions of particles interacting throught hard-sphere, Lennard-Jones and Yukawa (or Debye-H\"uckel) potentials \citep{Banetta_Yukawa}.
Furthermore, hydrodynamic interactions can be included as well, to some extent, in the above framework.

The pair correlation function obtained by solving the pair Smoluchowki equation only holds in very dilute conditions, where the $g (\mathbf{r}_{1} , \mathbf{r}_{2})$ function is not affected by the $N-2$ particles surrounding the two placed at $\mathbf{r}_{1}$ and $\mathbf{r}_{2},$ respectively. This limit does not hold when pair correlations are obtained by numerical simulations, as in Ref. \citep{Morris_Katyal}. 

A theoretical scheme able to compute the pair correlation function of a sheared colloidal suspension, at concentrated packing fractions as those considered in simulations, is missing. To fill this gap, in this paper we combine the analytical treatment introduced in Ref. \citep{Banetta_Zaccone_PRE} with the integral equation theories of the liquid-state \citep{Hansen_McDonald_BOOK}. While we introduce a theoretical method in principle suitable for any pair potential, we focus on the case of a hard-sphere colloidal suspension. This allows us to test our theoretical predictions with results of numerical simulations present in literature. We exploit the analytical solution obtained by following the method of Ref. \citep{Banetta_Zaccone_PRE} for hard spheres to build a potential of mean force $u_{\textrm{eff}},$ containing the effect of the flow field on the microscopic structure. Crucial to build the potential of mean force is to include hydrodynamic interactions, and treat them differently in the compressing and extensional sectors, respectively. We insert $u_{\textrm{eff}}$ in the Percus-Yevick relation and use the latter as a closure to solve the Ornstein-Zernike integral equation for a wide range of either the packing fraction $\eta$ and the P\'eclet number $\textrm{Pe}.$ As it is well-known, the OZ equation expresses the pair correlation function as a sum of a direct correlation between two particles, and the indirect correlation propagated via increasingly larger number of intermediate particles. It is then suitable to deal with suspensions in the more concentrated regime.
 
We obtain profiles for the correlation function which are in  very good agreement with  numerical results from Ref. \citep{Morris_Katyal} up to rather large values of $\eta,$ independently of the considered value of $\textrm{Pe}.$ We then extract scaling laws for the value of the pair correlation function at contact as a function of the $\textrm{Pe}$ number at fixed $\eta,$ and as a function of $\eta$ at fixed $\textrm{Pe}$ number. In the former case, we obtain a scaling law in agreement with the simulation study of Ref. \citep{Morris_Katyal}. In the latter case, we obtain a scaling law which may open the way for a non-equilibrium equation of state of strongly sheared liquids. Finally we employ our method to investigate the effect of the shear flow on the structure factor $ S (k)$ of the system. The theory predicts a consistent enhancement of $S (k)$ at $k \to 0,$ upon increasing the $\textrm{Pe}$ number. We argue this behavior to unveil the onset of a shear-induced transition from the isotropic to a non-uniform state, of the type discussed by Brazovskii \citep{Brazovski}.

The paper is organized as follows. In section II we introduce our theoretical scheme. In section III, we present our predictions. Finally, in section IV, we draw our conclusions.

\section{Theory}

\begin{figure}
\centering
\includegraphics[width = 0.8 \linewidth]{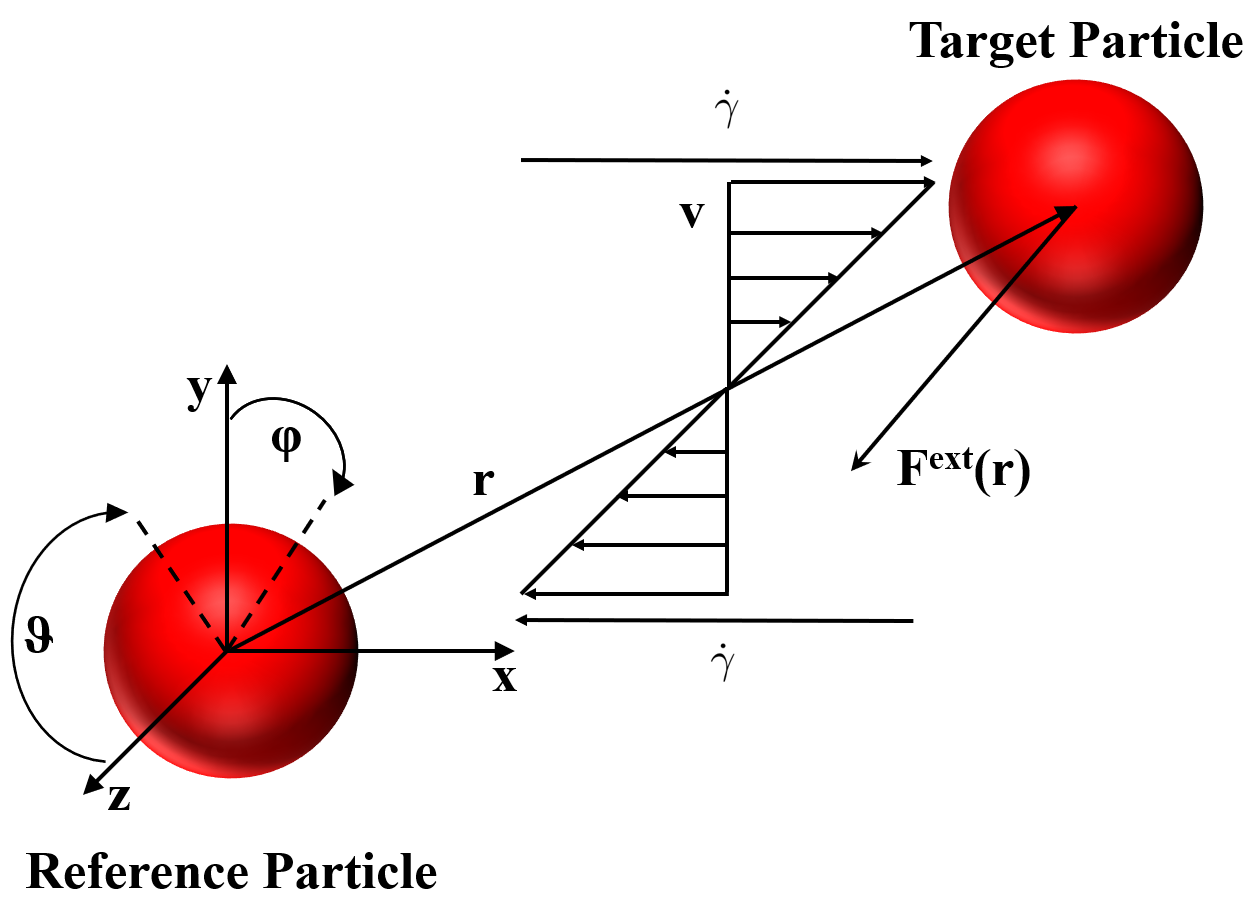}
 	\caption{Schematic illustration of a hard-sphere colloidal suspension subjected to a simple shear flow. In the dilute regime, the probability of finding a \textit{target} particle at a distance $\mathbf{r}$ from a \textit{reference} particle is weakly affected by the surrounding particles dispersed in the suspension. We consider a simple shear flow in the $x$-direction with its gradient in the $y$-direction such that, in the absence of hydrodynamic interactions, the fluid flow velocity at point $\mathbf{r}$ is given by $\boldsymbol{\kappa} \cdot \mathbf{r} = (\dot{\gamma} y, 0,0),$ with  $\boldsymbol{\kappa}$ and $\dot{\gamma}$ the velocity gradient tensor and the shear rate, respectively. }
	\label{fig:2_Body_System}
\end{figure}

As stated in the Introduction, the pair correlation function $g (\mathbf{r}_{1}, \mathbf{r}_{2})$ of a colloidal suspension at equilibrium describes the probability of finding a first particle in a volume $d \mathbf{r}$ centered at $\mathbf{r}_{1},$ and a second particle in a volume $d \mathbf{r}$ centered at $\mathbf{r}_{2},$ irrespective of the position of the remaining $N-2$ particles. For the sake of convenience, we will refer to the first and second particles as the \textit{reference} and the \textit{target} particles, respectively, throughout the paper. In case the colloidal particles have an isotropic spherical shape, the pair correlation function depends only on the relative distance $\mathbf{r} \equiv \mathbf{r}_{2} - \mathbf{r}_{1}$ between the particles, and in the presence of an external shear flow will also depend on time. As a consequence the pair correlation function can be indicated by $g (\mathbf{r}, t).$ The situation is shown in Fig. \ref{fig:2_Body_System}, where a Cartesian reference frame is introduced with origin at the center of the reference particle, i. e. $\mathbf{r}_{1}= (0,0,0).$ 

The temporal evolution of the $g (\mathbf{r}, t)$ function is given by the pair Smoluchowski equation with shear flow \citep{Dhont_BOOK,Brader_2010}
\begin{equation}
\begin{aligned}
\frac{\partial g (\mathbf{r},t)}{\partial t} &+ \boldsymbol{\nabla} \cdot \big[ \mathbf{v} (\mathbf{r}) g (\mathbf{r}, t) - \mathbf{D} (\mathbf{r}) \cdot \boldsymbol{\nabla}  g (\mathbf{r},t) \big] \\ 
&= - \boldsymbol{\nabla} \cdot \big[ \mathbf{D} (\mathbf{r}) \cdot \beta \mathbf{F} (\mathbf{r}) g (\mathbf{r}, t) \big],
\label{Pair_Smoluchowski}
\end{aligned}
\end{equation}
where $\mathbf{F} (\mathbf{r})$ describes the force acting between the colloidal particles in the suspension, $\mathbf{v} (\mathbf{r})$ is the relative velocity between the particles and $\mathbf{D} (\mathbf{r})$ is the \textit{diffusion tensor}. As is clear from Eq. \eqref{Pair_Smoluchowski}, the dynamics of the colloidal suspension is determined by competing effects of interparticle interaction, diffusion, and external flow.

We consider the dilute regime where triplet correlations can be safely neglected \citep{Brader_2010}. It follows that the $\mathbf{F} (\mathbf{r})$ term appearing in the previous equation describes the force acting between our chosen pair of particles due only to the direct potential interaction $u (r),$ i. e.
\begin{equation}
\mathbf{F} (\mathbf{r}) = - \boldsymbol{\nabla} u (r),
\end{equation} 
where $r \equiv |\mathbf{r}|.$ Observe that we only consider isotropic pair potentials that depend on the modulus of the relative distance $\mathbf{r}$ between the particles. When hydrodynamic interactions between the pair of particles are included in the theoretical treatment, the diffusion tensor $\mathbf{D} (\mathbf{r})$ present in Eq. \eqref{Pair_Smoluchowski} can be written as \citep{Brader_2010}
\begin{equation} \label{diffusion_tensor_hydro}
\mathbf{D} (\mathbf{r}) = 2 \mathcal{D}_{0} \bigg[  \frac{\mathbf{r} \mathbf{r}}{r^{2}} G(r) + \bigg( \boldsymbol{\delta} - \frac{\mathbf{r} \mathbf{r}}{r^{2}} \bigg) H(r) \bigg],
\end{equation}
where $\mathcal{D}_{0} \equiv k_{B} T /\big( 3 \pi \eta_{0} \sigma \big)$ is the diffusion coefficient of a single spherical particle of diameter $\sigma$ in a medium of viscosity $\eta_{0},$ $\mathbf{r} \mathbf{r}$ denotes the dyadic product, $\boldsymbol{\delta}$ is the identity matrix, and $G(r)$ and $H(r)$ are scalar functions containing the details of the hydrodynamic interactions. Finally, the relative velocity of the particles is given by \citep{Brader_2010}
\begin{equation}
\mathbf{v} (\mathbf{r}) = \boldsymbol{\kappa} \cdot \mathbf{r} + \mathbf{C} (\mathbf{r}) : \bar{\boldsymbol{\kappa}},
\end{equation}
where $\boldsymbol{\kappa}$ is the velocity gradient tensor, $\boldsymbol{\kappa}^{T}$ its transpose and $\bar{\boldsymbol{\kappa}} \equiv ( \boldsymbol{\kappa} + \boldsymbol{\kappa}^{T} )/2.$ The third-rank tensor $\mathbf{C} (\mathbf{r})$ is known as the \textit{hydrodynamic resistance} tensor. While $\boldsymbol{\kappa} \cdot \mathbf{r}$ describes the motion of the fluid suspension because of the applied shear, the term $\mathbf{C} (\mathbf{r}) : \bar{\boldsymbol{\kappa}}$ describes the disturbance of the affine flow due to the presence of the particles. The latter term can be written as
\begin{equation} \label{C_tensor}
\mathbf{C} (\mathbf{r}) : \bar{\boldsymbol{\kappa}} = - r \bigg[ \frac{\mathbf{r} \mathbf{r} \cdot \bar{\boldsymbol{\kappa}} \cdot \mathbf{r}}{r^{3}} A(r) + \bigg( \boldsymbol{\delta} - \frac{\mathbf{r} \mathbf{r}}{r^{2}} \bigg) \cdot \frac{\bar{\boldsymbol{\kappa}} \cdot \mathbf{r}}{r} B(r) \bigg],
\end{equation}
where $A(r)$ and $B(r)$ are to be determined. 

Neglection of hydrodynamic interactions in the introduced theoretical framework can be obtained by imposing $G(r)=H(r)=1$ and $A(r)=B(r)=0$ in Eqs. \eqref{diffusion_tensor_hydro} and  \eqref{C_tensor}, respectively. In this case the diffusion tensor and the relative velocity between the particles reduce to $\mathbf{D} (\mathbf{r}) = 2 \mathcal{D}_{0}  \boldsymbol{\delta}$ and  $\mathbf{v} (\mathbf{r}) = \boldsymbol{\kappa} \cdot \mathbf{r},$ respectively. 

In this paper we consider suspensions under the action of a simple shear flow directed along the $x$-axis with gradient along the $y$-axis, as shown in Fig. \ref{fig:2_Body_System}. Thus, by indicating with $\dot{\gamma}$ the shear rate, the velocity gradient tensor $\boldsymbol{\kappa}$ reads
\begin{equation}
\begin{aligned} 
\boldsymbol{\kappa} &=  \begin{pmatrix} 0 & \dot{\gamma} & 0 \\ 0 &
0 & 0 \\ 0 & 0 & 0
\end{pmatrix},
\end{aligned}
\end{equation}
from which $\boldsymbol{\kappa} \cdot \mathbf{r} = (\dot{\gamma} y, 0, 0).$ Furthermore, throughout the paper we will only consider steady-state situations where $\partial g(\mathbf{r}, t) / \partial t =0.$ As a consequence  we will neglect the time dependence of the  pair correlation function and indicate the latter by ``simply" $ g (\mathbf{r}).$

\begin{figure}
\centering
	\includegraphics[width = 0.9 \linewidth]{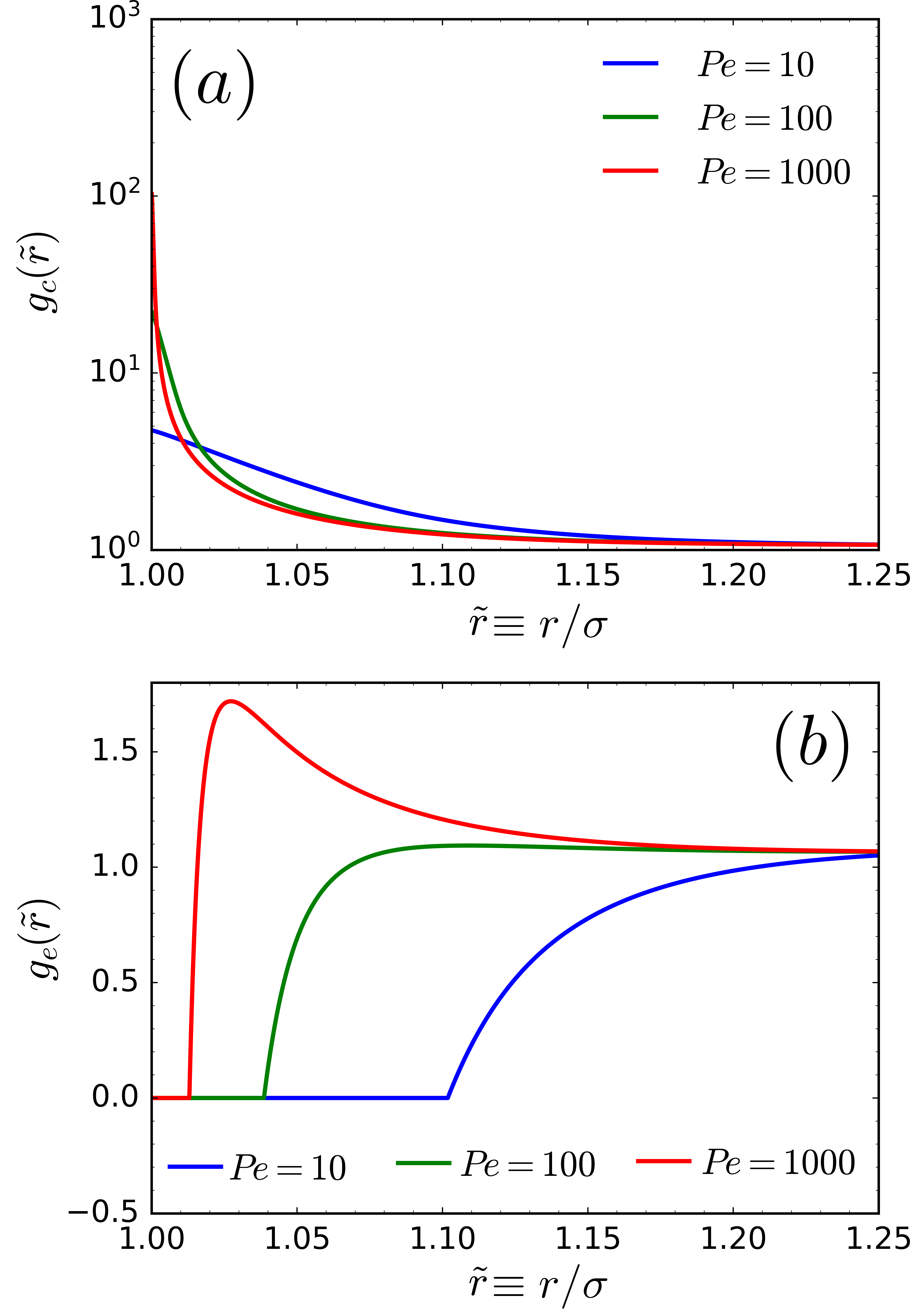}
	\caption{Angular average of the pair correlation function of a hard-sphere colloidal suspension under shear flow, over the compressing ($a$) and extensional ($b$) sectors of the solid angle, respectively. In the compressing sectors, the averaged pair correlation function shows a peak near contact, which increases with the $\textrm{Pe}$ number. The extensional sectors, feature a depletion layer near contact where the pair correlation function is identically null. Either in (a) and in (b), the values of $\tilde{r}$ for which the pair correlation function is different from $1$ decreases upon increasing of $\textrm{Pe}$ number.}
	\label{pmf}
\end{figure}

\begin{figure}
	\centering
		\includegraphics[width = 0.9 \linewidth]{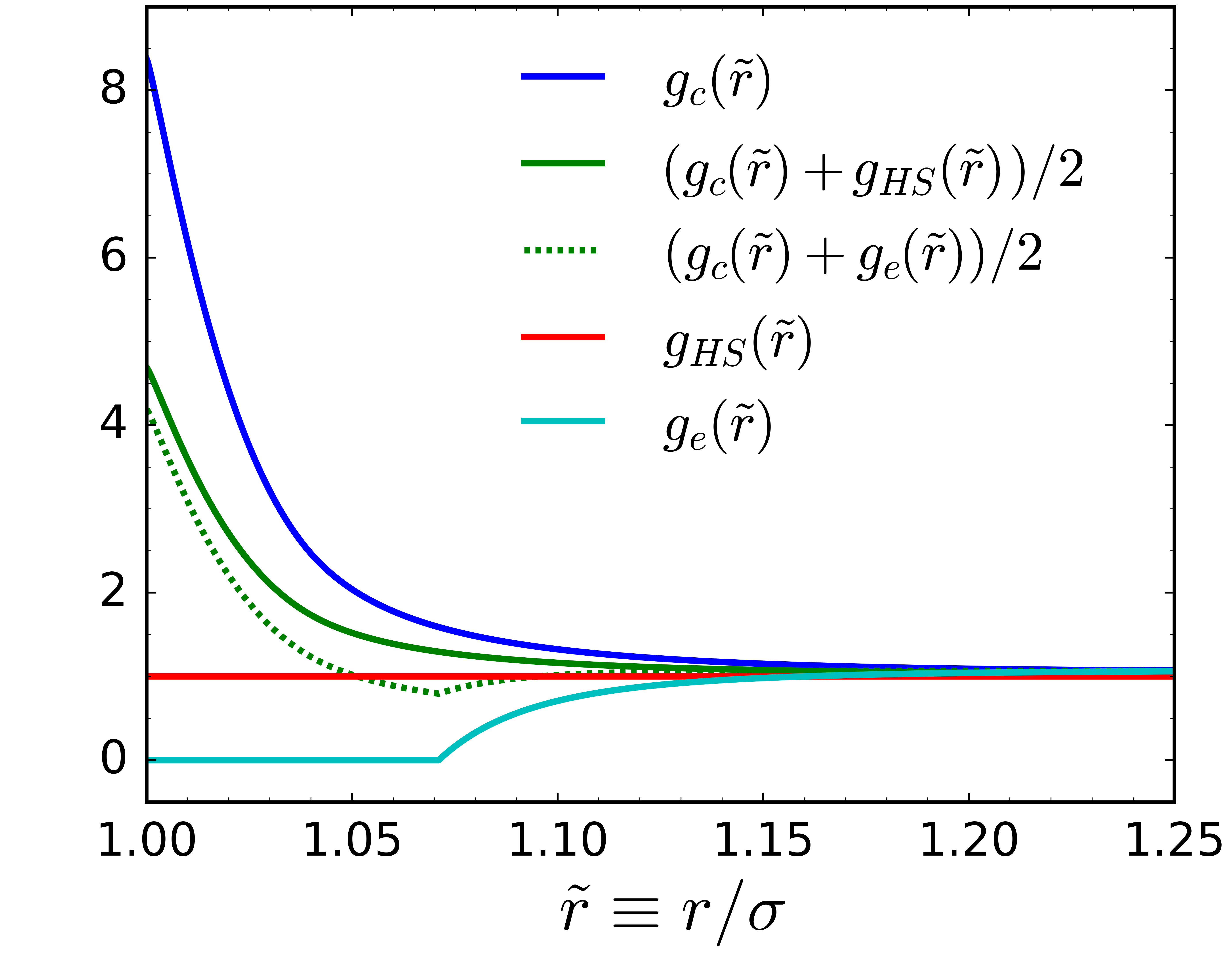}
	\caption{Angular average of the pair correlation function of a hard-sphere colloidal suspension under shear flow, over the compressing (blue) and extensional (light blue) sectors of the solid angle, respectively, at fixed $\textrm{Pe}=25.$ The dashed green line indicates the linear combination of $g_{c} (\tilde{r})$ and $g_{e} (\tilde{r})$ obtained from Eq. \eqref{average} of the main text. The full green line shows the linear combination of $g_{c} (\tilde{r})$ and $g_{e} (\tilde{r})$ obtained from Eq. \eqref{average}, when $g_{e} (\tilde{r}) \equiv g_{\textrm{HS}} (\tilde{r}).$ The function $g_{\textrm{HS}} (\tilde{r})$ (see red line) represents the pair correlation function of a hard-sphere gas (an extremely dilute suspension) in the absence of any shear flow. }
	\label{pmf1}
\end{figure}

Recently, a new scheme to solve analytically the pair Smoluchowski equation \eqref{Pair_Smoluchowski} in steady-state was proposed in Ref. \citep{Banetta_Zaccone_PRE}. The method holds for a generic pair potential $u (r)$ and takes into account the hydrodynamic interactions by considering approximations for the $G(r),$ $H(r),$ $A(r)$ and $B(r)$ functions introduced above. The starting point of the new strategy is to realize that Eq. \eqref{Pair_Smoluchowski} is a partial differential equation whose solution is typically challenging even numerically. To overcome this difficulty, it is then proposed to consider an angular average of the pair Smoluchowski equation, thus replacing Eq. \eqref{Pair_Smoluchowski} with an effective ordinary differential equation for the orientation averaged pair correlation function, $g (r).$ 

However, as already noticed in Ref. \citep{Banetta_Zaccone_PRE}, when Eq. \eqref{Pair_Smoluchowski} is averaged over the full solid angle $\Omega \equiv (\theta, \phi)$ with $\theta \in [0,\pi]$ and $\phi \in [0, 2 \pi]$, a vanishing net effect of the shear flow on the $g (\mathbf{r})$ function results. To solve this problem, the authors of Ref. \citep{Banetta_Zaccone_PRE} observed that a generic shear flow can be divided into different \textit{sectors} of the solid angle $\Omega,$ on the basis of the sign of the radial component of the relative velocity between the two particles, $v_{r} (\mathbf{r}).$ 

In the \textit{compressing} sectors $v_{r} (\mathbf{r})<0$ and the particles are pushed towards each other by the shear flow. By contrast, in the \textit{extensional} sectors $v_{r} (\mathbf{r})>0$ and the particles are pulled away from one another by the shear flow. As showed in Ref. \citep{Banetta_Zaccone_PRE}, the compressing sectors are identified by the angles $\theta_{c} \in [0, \pi],$ $\phi_{c}  \in [\pi/2, \pi] $ and $\phi_{c}  \in [3 \pi /2, 2 	\pi],$ while the extensional sectors are identified by the angles $\theta_{e}  \in [0, \pi],$ $\phi_{e} \in [0, \pi/2]$ and $\phi_{e} \in [\pi, 3 \pi /2].$ Two ordinary differential equations result from the angular average of Eq. \eqref{Pair_Smoluchowski} over the compressing and extensional sectors of $\Omega,$ respectively, which can be solved independently for several values of the $\textrm{Pe}$ number by means of the so-called \textit{intermediate asymptotics methodology} \citep{Bender_Orszag}.
Two distinct functions are hence obtained as output of the analytical treatment: a $g_{c} (r)$ function describing the average of the pair correlation function $g (\mathbf{r})$ over the compressing sectors, and a $g_{e} (r)$ function describing the average of the pair correlation function $g (\mathbf{r})$ over the extensional sectors. An estimate of the average of the pair correlation function over the full solid angle $\Omega$ can be finally obtained by combining the $g_{c} (r)$ and $g_{e} (r)$ functions. To remark that the validity of the obtained pair correlation function is limited to the very dilute regime $\eta \to 0,$ we indicate it as  $g_\text{0} (r)  \equiv g(r, \eta \to 0 ).$ We write
\begin{equation} \label{average}
    g_\text{0} (r)  \equiv \frac{g_\text{c} (r)  + g_\text{e} (r)}{2} ,
\end{equation}
where, as shown in Appendix A, $g_\text{0} (r)$ corresonds to the average of the pair correlation function $g (\mathbf{r})$ over the full solid angle $\Omega.$ 

We observe that $g_\text{0} (r)$ only depends on the modulus of the distance $\mathbf{r}$ between the particles, since the procedure of angular averaging comes at expense of loosing angular resolution. The introduced procedure, however, is of remarkable importance since it considerably simplifies the pair Smoluchowski equation and allows us to solve it analytically.

As noticed, a vanishing net effect of the shear flow on the pair distribution function results when Eq. \eqref{Pair_Smoluchowski} is averaged over the full solid angle $\Omega.$ As a consequence, in this case, the solution of the resulting steady-state effective Smoluchowski equation is a pair correlation function always identical to 1, independently of the $\textrm{Pe}$ number. In other words, the positions of the reference and the target particles are always independent of each other. This is not the case, however, for the $g_{0} (r)$ function obtained by combining the $g_{c} (r)$ and $g_{e} (r)$ solutions as in Eq. \eqref{average}. As it will be discussed in the next subsection, this is due to a proper treatment of hydrodynamic interactions in the different sectors of the solid angle.

The purpose of this paper is to extend the range of validity of the  $g_{0} (r)$ function \eqref{average} at larger values of the packing fraction $\eta,$ for the particular case of a hard-sphere colloidal suspension under shear flow. In this region of $\eta,$ the effect on the pair correlation function of the $N-2$ particles surrounding the reference and target particles cannot be neglected. We show this effect can be taken into account when the analytical solution of Eq. \eqref{Pair_Smoluchowski} proposed in Refs. \citep{Banetta_Zaccone_PRE} is combined with the well-known integral equation theory of liquids \citep{Hansen_McDonald_BOOK}. To distinguish the pair correlation function holding in this larger range of the packing fraction from the $g_{0}(r)$ holding in the very dilute limit, we will indicate the former by $g(r)$ throughout the paper.

A cornerstone of liquid-state theory is the so-called Ornstein-Zernike (OZ) integral equation which, for a homogeneous and isotropic system, is given by \citep{Hansen_McDonald_BOOK}
\begin{equation} \label{OZ}
h (r) = c(r) + \rho \int_{V} d \mathbf{r}_{3} \ c \big( | \mathbf{r}_{1} - \mathbf{r}_{3} | \big) h \big( | \mathbf{r}_{3} - \mathbf{r}_{2} | \big), 
\end{equation}
where $r \equiv |\mathbf{r}_{1} - \mathbf{r}_{2}|,$ $h (r)$ are $c (r)$ are the \textit{total} and the \textit{direct} correlation functions, respectively, while $\rho$ is the number density. 
The OZ equation expresses the $g(r)$ function as a sum of the \textit{direct} correlation function between the reference and target particles, and the \textit{indirect} correlation propagated via increasingly larger number of intermediate particles. It is then suitable to deal with suspensions in the more concentrated regime. To find the $g (r),$ the OZ equation has to be supplemented by an independent closure relation between $c(r),$ $h(r),$ and the pair potential $u (r).$ In this paper we close the OZ equation with the so-called Percus-Yevick relation \citep{caccamoIET}
\begin{equation} \label{PY}
c(r) = g(r) - g(r) e^{\beta u (r)},
\end{equation}
which has been proved to be accurate for hard-sphere systems \citep{Mulero}. Eqs. \eqref{OZ} and \eqref{PY} form a self-closed system which, for a fixed pair potential $u (r),$ can be solved in order to find $g(r).$ 

However, the dependence on the $\textrm{Pe}$ number, and hence the effect of the shear flow, is absent either in Eq. \eqref{OZ} and in Eq. \eqref{PY}. We here exploit the solution $g_{0} (r)$ of the pair Smoluchowski equation to define a potential of mean force $u_{\textrm{eff}} (r, \textrm{Pe})$ which contains the effect of the flow field on the microscopic structure. More precisely we define
\begin{equation} \label{effectiv_pp}
    \beta u_{\textrm{eff}} (r,\textrm{Pe}) \equiv \begin{cases}
     \infty \quad r < \sigma \\
     - \log \big[ g_{0} (r) \big]  \quad r \ge \sigma 
    \end{cases},
\end{equation}
where $g_{0} (r)$ is given by Eq. \eqref{average}. The introduced pair potential has a clear dependence on the $\textrm{Pe}$ number. By inserting $u_{\textrm{eff}} (r, \textrm{Pe})$ in the PY closure and solving the resulting equation together with the OZ equation \eqref{OZ}, allows us to obtain a $g (r)$ function which depends, at the same time, on the $\textrm{Pe}$ number and contains contributions from surrounding particles. This scheme, hence, allows us to investigate the microscopic structure in a range of $\eta,$ so far unexplored by means of theoretical methods.

We notice that a simple shear flow is a non-conservative external field for which a potential of mean force, in principle, does not exist. Using Eq. \eqref{effectiv_pp} in our theoretical scheme is then an approximation.

Finally, it is very important to notice that in the outlined framework, the OZ equation is used to determine the pair correlation function of an out-of-equilibrium system, as is a colloidal suspension under shear flow. This may be a rather disputable assumption, since the OZ equation has been typically employed for equilibrium systems. Addressing this point at the theoretical level, however, is beyond the scope of this paper. We here limit ourselves to verify the validity of our method \textit{a posteriori}, by a systematic comparison of our theoretical predictions with simulations data present in the literature.

\begin{figure*}
	\centering
	\includegraphics[width = 0.9 \linewidth]{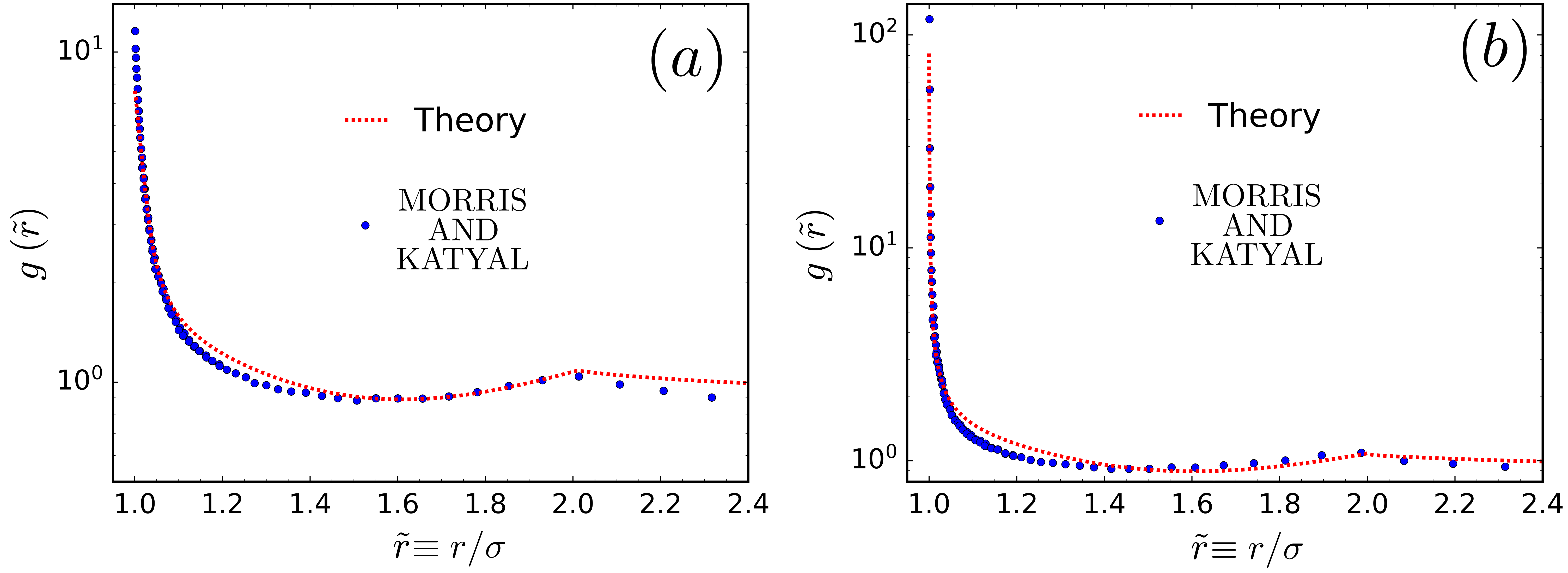}
 	\caption{Pair correlation function of a hard-sphere colloidal suspension under shear flow at packing fraction $\eta=0.30,$ and $\textrm{Pe}= 25$ (a) and $\textrm{Pe}= 1000$ (b), respectively. Dashed red lines represent results from our theoretical scheme, while points are results from numerical simulations of Ref. \citep{Morris_Katyal}. A very good agreement between predictions of theory and results from numerical simulations can be observed for both values of the $\textrm{Pe}$ number. }
	\label{comparison_eta035}
\end{figure*}

\begin{figure*}
	\centering
	\includegraphics[width = 0.9 \linewidth]{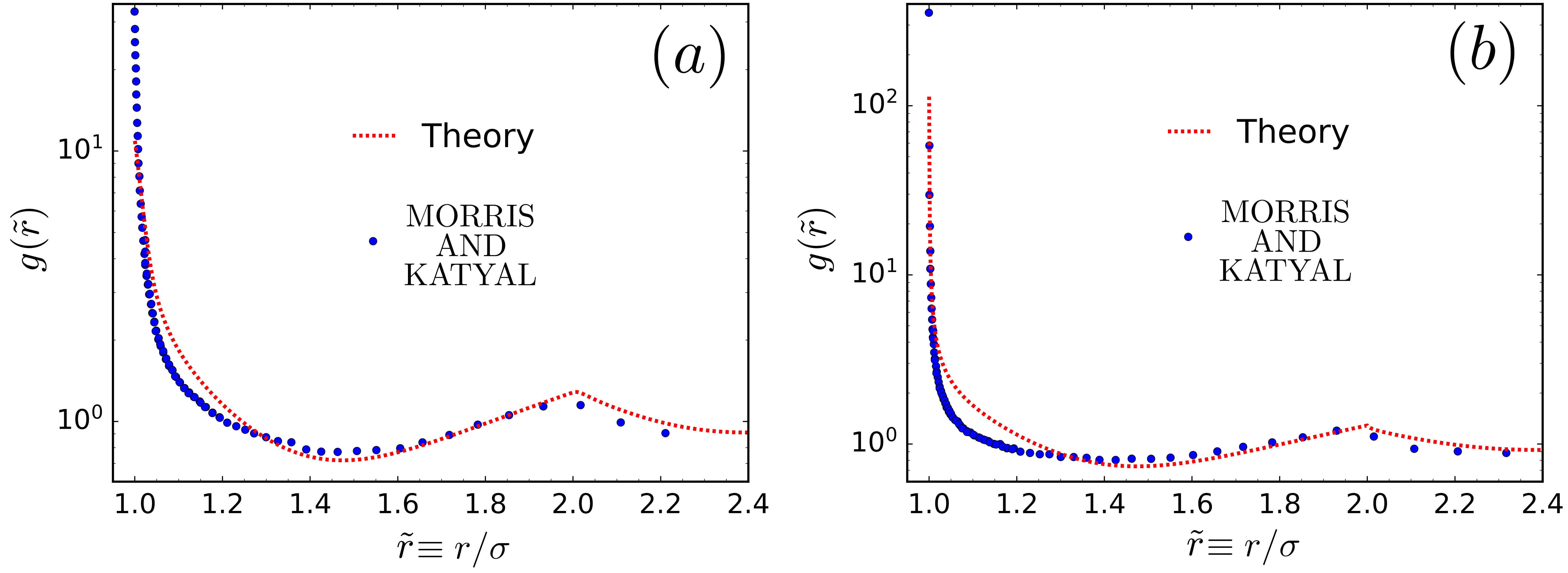}
 	\caption{Same as in Fig. \ref{comparison_eta035}, but for packing fraction $\eta=0.45.$ While a good qualitative agreement between theoretical predictions and numerical findings is still obtained, a worse quantitative agreement than in Fig. \ref{comparison_eta035} can be observed.} 
	\label{comparison_eta040}
\end{figure*}

\subsection{Potential of mean force for hard spheres under shear flow}

\begin{figure*}
\centering
\includegraphics[width = 0.9 \linewidth]{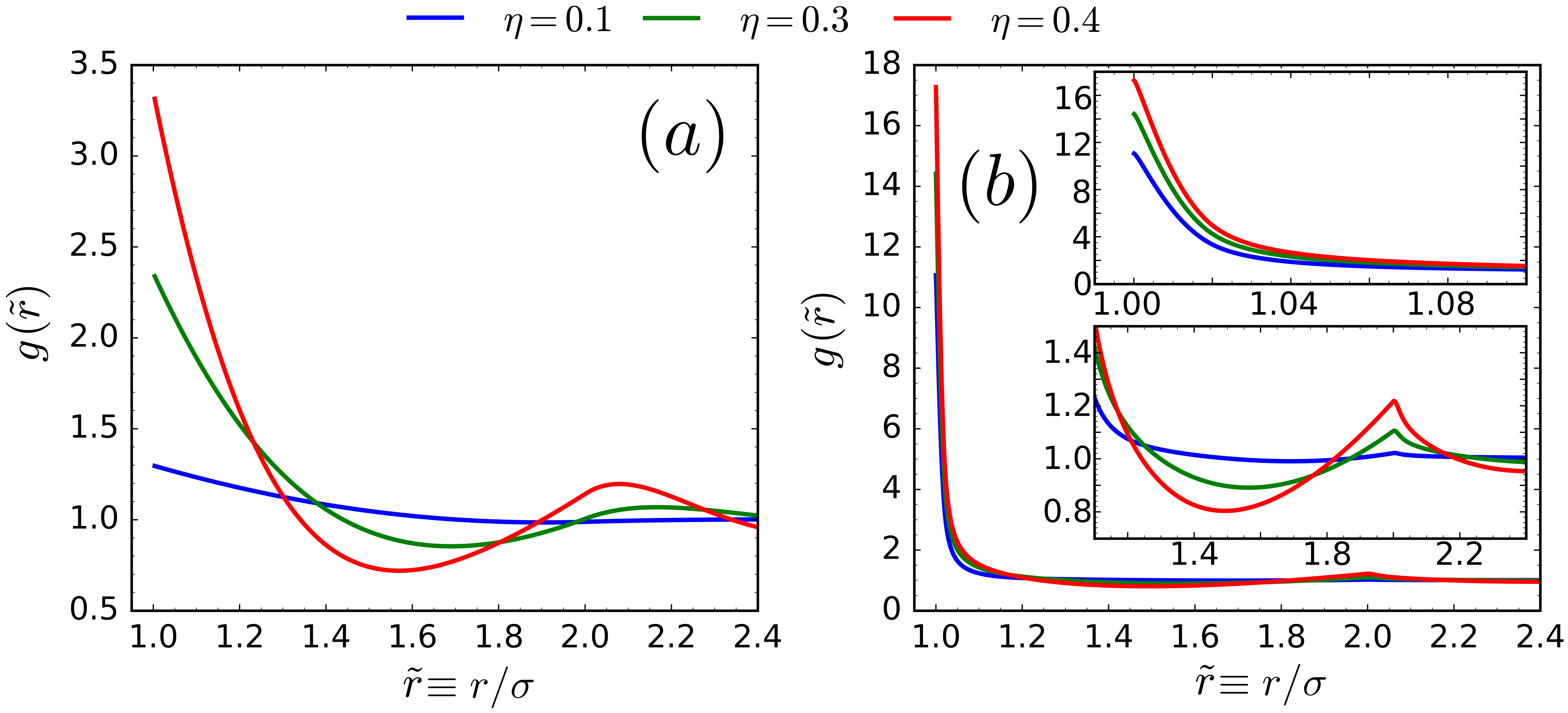}
	\caption{Pair correlation function of a hard-sphere colloidal suspension for several values of the packing fraction $\eta.$ In (a) $\textrm{Pe}=0,$ i. e. the shear flow is absent, while in (b) $\textrm{Pe}=50.$ The insets in (b) show a zoom in the regions $\tilde{r} \in [1.0,1.1]$ and $\tilde{r} \in [1.1, 2.4],$ respectively. When $\textrm{Pe}=50,$ the value at contact of the pair correlation function is much larger than the same value in case $\textrm{Pe}=0,$ for all the considered values of $\eta.$ Moreover, while in (a) the position of the second peak varies with $\eta,$ in (b) this peak is always located at $\tilde{r}=2.$}
	\label{structure_Pe}
\end{figure*}

We here show how to build the potential of mean force Eq. \eqref{effectiv_pp} in the case of a hard-sphere colloidal suspension under shear flow. 

As it is well-known, the hard-sphere pair potential is
\begin{equation} \label{potential_HS}
   \beta u_{\textrm{HS}}(r)= \begin{cases}
     \infty \quad r < \sigma \\
     0 \ \ \ \ \quad r \ge \sigma
    \end{cases},
\end{equation}
where $\sigma$ is the particle diameter. For this system, the steady-state pair Smoluchowski equation was solved in Ref. \citep{Banetta_Zaccone_PRE} by means of the intermediate asymptotics methodology. Here we follow the treatment introduced in that paper. We start by discussing the approximations considered to include the hydrodynamic interactions in our theoretical framework. First of all, we model the microscopic diffusion matrix $\mathbf{D} (\mathbf{r})$ present in Eq. \eqref{Pair_Smoluchowski}, and given by Eq. \eqref{diffusion_tensor_hydro}. We introduce a spherical reference system with origin at the center of the reference particle (see Fig. \ref{fig:2_Body_System}). We assume all the off-diagonal elements of $\mathbf{D} (\mathbf{r})$ to be null, i. e. $D_{ij} (\mathbf{r}) =0$ for $i,j=1,2,3$ and $i \neq j.$ Furthermore we assume $H (r)=0$ such that $D_{22} (\mathbf{r}) = D_{33} (\mathbf{r})=0$ and the only non-null element of the diffusion tensor is $D_{11} (r) = 2 \mathcal{D}_{0} G (r).$ $G(r)$ is the hydrodynamic function for the viscous retardation (also known as lubrication effect), and the chosen form for the diffusion tensor $\mathbf{D} (\mathbf{r})$ is equivalent to assume the viscous retardation to act only radially between the particles. As noticed in Ref. \citep{Banetta_Zaccone_PRE}, the $G(r)$ function cannot have the same functional form in either the compressional and extensional sectors of the solid angle. $G(r),$ indeed, describes a repulsive effect experienced by particles approaching each other radially, due to the squeezing of the liquid between them. It then plays a relevant role in the compressing sectors, while it is negligible in extensional sectors. We model $G (r)$ in the compressing sector through a polynomial fit to the rigorous solution to the Stokes equation for the specific case of two particles approaching each other \citep{Banetta_Zaccone_PRE}
\begin{equation} \label{G_lubr_compressing}
G_{c} (r) = \frac{6 h^{2} + 4 h}{6 h^{2} + 13 h +2},
\end{equation}
where $h \equiv r -\sigma$ is the surface distance between the particles. By contrast, we assume the lubrication force to be negligible in the extensional sectors, by imposing
\begin{equation} \label{G_lubr_extensional}
G_{e} (r) =1.
\end{equation}

As discussed, the hydrodynamic functions also enter the expression for the relative velocity $\mathbf{v} (\mathbf{r})$ between the particles (see Eq. \eqref{C_tensor}). As shown in Refs. \citep{Banetta_Zaccone_PRE, adler}, $\mathbf{v} (\mathbf{r})$ can be written as 
\begin{equation} \label{velocity_HS}
    \begin{aligned}
&        v_{r} (r, \theta, \phi) = \dot{\gamma} r \big[ 1 - A(r) \big] \sin^{2} \theta \sin \phi \cos \phi, \\
&        v_{\theta} (r, \theta, \phi) = \dot{\gamma} r \big[ 1 - B(r) \big] \sin \theta \cos \theta \sin \phi \cos \phi,  \\
&         v_{\phi} (r, \theta, \phi) = \dot{\gamma} r \sin \theta \bigg[ \cos^{2} \phi - \frac{B(r)}{2} \cos 2 \phi \bigg] .
    \end{aligned}
\end{equation}
To model the hydrodynamic functions we follow Refs. \citep{Melis,Banetta_Zaccone_PRE,Banetta_Yukawa}. We assume $A (r)$ to be given by 
\begin{equation}
\begin{aligned} 
A (r) &= \frac{A_{1}}{\big( 2 r \big)^{5}} + \frac{A_{2}}{\big( 2 r \big)^{6}}  - \frac{A_{3}}{\big( 2 r  \big)^{7}} + \frac{A_{4}}{\big( 2 r  \big)^{8}}
\end{aligned}
\end{equation}
where $A_{1} = 113.2568894, \ A_{2} =307.8264828, \ A_{3}=2607.54064288$ and $A_{4}=3333.72020041,$ and $B (r)$ to be given by 
\begin{equation}
\begin{aligned} 
B (r) &= \frac{B_{1}}{\big( 2 r  - B_{2} \big)^{\beta_{1}}} - \frac{B_{3}}{\big( 2 r  - B_{4} \big)^{\beta_{2}}}
\end{aligned}
\end{equation}
where $B_{1} = 0.96337157, \ B_{2} =1.90461683, \ B_{3}=0.93850774, \ B_{4}=1.90378420, \ \beta_{1}=-1.99517070$ and $\beta_{2}= 2.01254004.$

Having specified the approximations for the hydrodynamic functions, we need to solve the pair Smoluchowski equation \eqref{Pair_Smoluchowski}. To simplify the calculation, we first introduce the dimensionless quantities
\begin{equation}
    \begin{aligned}
&        \tilde{\textbf{r}} \equiv \textbf{r}/\sigma, \\
&        \tilde{\boldsymbol{\nabla}} \equiv \boldsymbol{\nabla} \sigma, \\
&        \tilde{u} \equiv \beta u  , \\
    \end{aligned}
\end{equation}
and $\tilde{\textbf{v}} \equiv \textbf{v}/(\sigma \dot{\gamma}).$ Using these, in steady-state, Eq. \eqref{Pair_Smoluchowski} becomes
\begin{equation}
     \dfrac{1}{\text{Pe}}\biggl[ G (\tilde{r}) \biggl( \tilde{\boldsymbol{\nabla}} + \tilde{\boldsymbol{\nabla}} \tilde{u}(\tilde{r}) \biggr) g(\tilde{\textbf{r}}) \biggr] = \tilde{\textbf{v}}(\tilde{\textbf{r}}) g(\tilde{\textbf{r}}),
    \label{Diluted_2_Body_Smoluchowski_Equation}
\end{equation}
where we have used the definition Eq. \eqref{definition_Peclet} of the $\textrm{Pe}$ number.

Following Ref. \citep{Banetta_Zaccone_PRE} , we average Eq. \eqref{Diluted_2_Body_Smoluchowski_Equation} over the compressing and extensional sectors of the solid angle. In the former case, we obtain a ordinary differential equation for the average of $g (\tilde{\mathbf{r}})$ over the compressing sectors, $g_{c} (\tilde{r}),$ in the latter case we obtain a ordinary differential equation for the average of $g (\tilde{\mathbf{r}})$ over the extensional sectors, $g_{e} (\tilde{r}).$ We solve the resulting equations perturbatively. To this aim we introduce a small perturbation parameter $\epsilon$ defined as the inverse of the $\textrm{Pe}$ number, i.e.
\begin{equation}
\epsilon \equiv 1 / \textrm{Pe}.
\end{equation}
The approach followed in Ref. \citep{Banetta_Zaccone_PRE} hence consists of the evaluation of two different power series related to two different regions of the radial coordinate domain: the \textit{outer} layer (far away from the reference particle), where the solution is slowly changing with $\tilde{r},$ and the \textit{boundary} layer (close to the reference particle), where the solution is steeply and very rapidly changing with $\tilde{r}.$ Details of the mathematical solution are presented in Appendix B. It is important to observe that, being based on an expansion in terms of $1 / \textrm{Pe},$ the analytical method holds mainly for large values of $\textrm{Pe},$ i. e. for strong shear flows.

We plot the obtained $g_{c} (\tilde{r})$ and $g_{e} (\tilde{r})$ functions in Fig. \ref{pmf}, for several values of the $\textrm{Pe}$ number. The behaviour of the pair correlation function in the compressing $\big($see Fig. \ref{pmf} $(a)  \big)$ and the extensional $\big($see Fig. \ref{pmf} $(b) \big)$ sectors is very different. In the compressing sectors, $g_{c} (\tilde{r})$ shows a peak at $\tilde{r}=1$ which increases with the $\textrm{Pe}$ number. By increasing $\textrm{Pe},$ indeed, the shear flow dominates over the hydrodynamic interactions which are instead repulsive. The extensional quadrants, on the other hand, feature a depletion layer near contact where the pair correlation function is identically null. This depletion layer is due to the presence of the hydrodynamic interactions and would disappear in case  $A(\tilde{r}) = B (\tilde{r}) = 0.$ In the latter case, if also $G_{c} (\tilde{r}) =1,$ we would find $g_{c} (\tilde{r}) =  2 - g_{e} (\tilde{r}),$ from which the $g_{0} (\tilde{r})$ function given by Eq. \eqref{average} would be identically equal to 1. It follows that, to investigate the effect of the shear flow, it is crucial to include the hydrodynamic interactions in our framework and to treat them properly in the compressing and extensional sectors of the solid angle, respectively.

As mentioned above, when $g_{c} (\tilde{r})$ and $g_{e} (\tilde{r})$ are known for a certain value of the $\textrm{Pe}$ number, a potential of mean force $u_{\textrm{eff}}$ can be built through Eq. \eqref{effectiv_pp}. To this aim, the function $g_{0} (\tilde{r})$ combining $g_{c} (\tilde{r})$ and $g_{e} (\tilde{r})$ needs to be considered, through Eq. \eqref{average}. For illustrative purposes, we show  $g_{c} (\tilde{r}),$ $g_{e} (\tilde{r})$ and $g_{0} (\tilde{r})$ at fixed $\textrm{Pe}=25$ in Fig. \ref{pmf1}. We observe (see dashed green line) that $g_{0} (\tilde{r})$ features an  unphysical kink. Our (approximate) treatment, indeed, does not guarantee $g_{0} (\tilde{r})$ to be  continuous in the first derivative. To solve this problem, we here neglect the depletion layer featured by the  $g_{e} (\tilde{r})$ function, and assume the latter to be identically equal to unity. In other words, we assume $g_{e} (\tilde{r}) \equiv g_{\textrm{HS}} (\tilde{r}),$ where $g_{\textrm{HS}} (\tilde{r})$ is the pair correlation function of a hard-sphere system in the very dilute regime, i. e. a hard-sphere gas, in the absence of any external flow. As it is known, while $g_{\textrm{HS}} (\tilde{r})=0$ for $\tilde{r} < 1,$ $g_{\textrm{HS}} (\tilde{r})=1$ for $\tilde{r} > 1.$ This approximation is justified by the lower weight of the depletion layer of the $g_{e} (\tilde{r})$ function in the average \eqref{average} with respect to the large peak characterizing the  $g_{c} (\tilde{r})$ function. Moreover, the amplitude of the depletion layer reduces by increasing the $\textrm{Pe}$ number. The $g_{0} (\tilde{r})$ obtained from Eq. \eqref{average} when $g_{e} (\tilde{r}) \equiv g_{\textrm{HS}} (\tilde{r})$ is plotted with full green line in Fig. \ref{pmf1}. As it is clear, the unphysical kink is not observed in this case and a smooth approximate solution is produced.

\subsection{Strategy recap}

We here summarize the strategy proposed in this paper to compute the pair correlation function $g (\tilde{r})$ for a system of concentrated hard spheres under shear flow.

For a fixed value of the $\textrm{Pe}$ number, we first insert the hard-sphere pair potential Eq. \eqref{potential_HS} into the pair Smoluchowski equation, Eq. \eqref{Diluted_2_Body_Smoluchowski_Equation}, and solve the latter by using the method introduced in Ref. \citep{Banetta_Zaccone_PRE} and briefly recalled in the previous section. In this way we find the compressional and extensional pair correlation functions $g_{c} (\tilde{r})$ and $g_{e} (\tilde{r}),$ respectively, which allow us to build the potential of mean force $u_{\textrm{eff}}$ according to Eq. \eqref{effectiv_pp}. We then insert $u_{\textrm{eff}}$ in the PY closure and solve the coupled OZ and PY equations, iteratively by means of the Picard's algorithm, for several values of the packing fraction $\eta.$ We repeat this scheme for different values of the $\textrm{Pe}$ number. 

\section{Results}

\begin{figure}
\centering
	\includegraphics[width = 0.9 \linewidth]{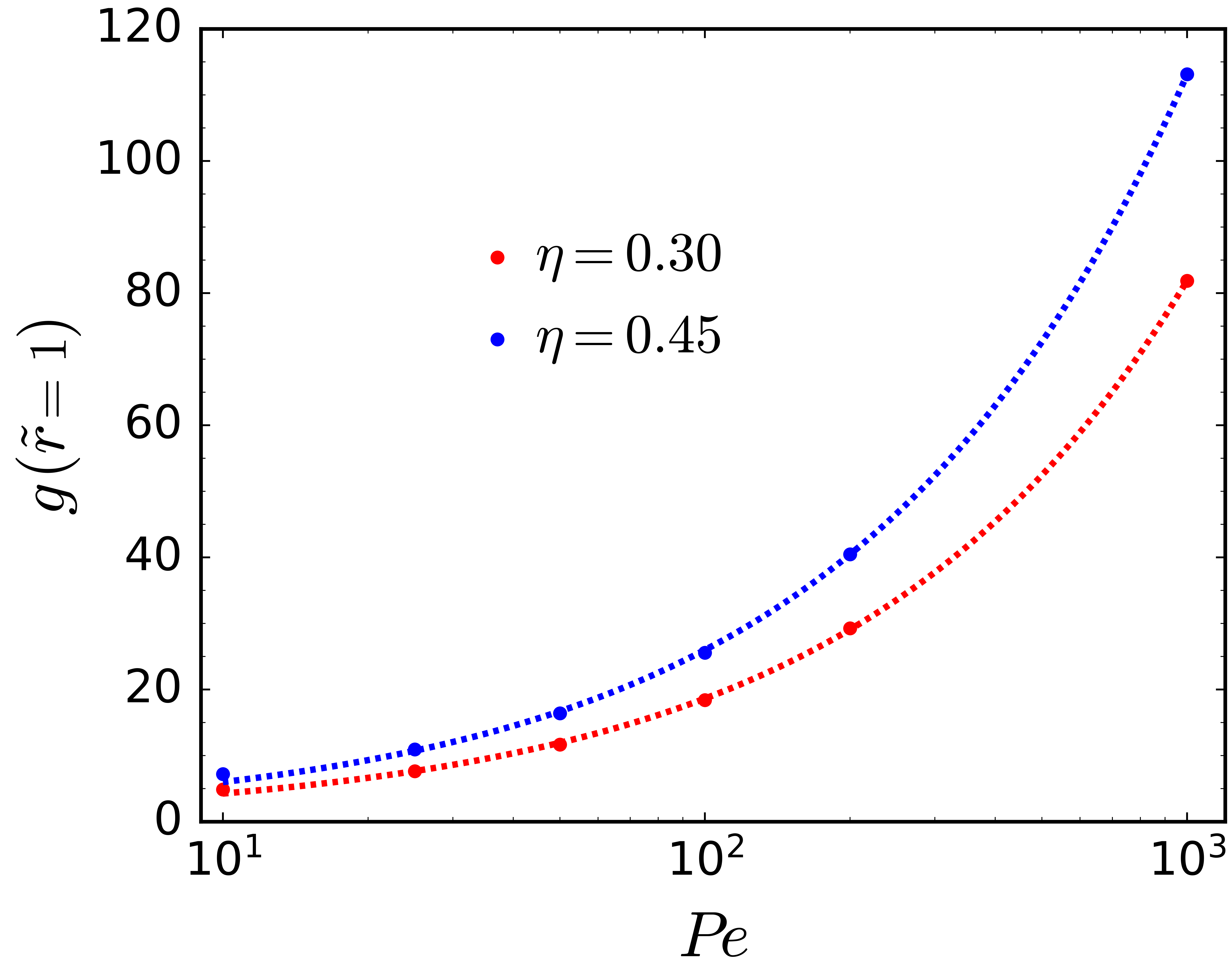}
	\caption{Value at contact of the pair correlation function as a function of the $\textrm{Pe}$ number, at fixed $\eta=0.30$ (red) and $\eta=0.45$ (blue), respectively. While points represent results of the introduced theoretical scheme, lines indicate a fit to Eq. \eqref{fit_Pe} of the main text. While in (a) $\alpha=0.96$ and $\beta=0.64,$ in (b) $\alpha=1.37$ and $\beta=0.64.$}
	\label{scaling_Pe}
\end{figure}

\begin{figure}
	\centering
	\includegraphics[width = 0.9 \linewidth]{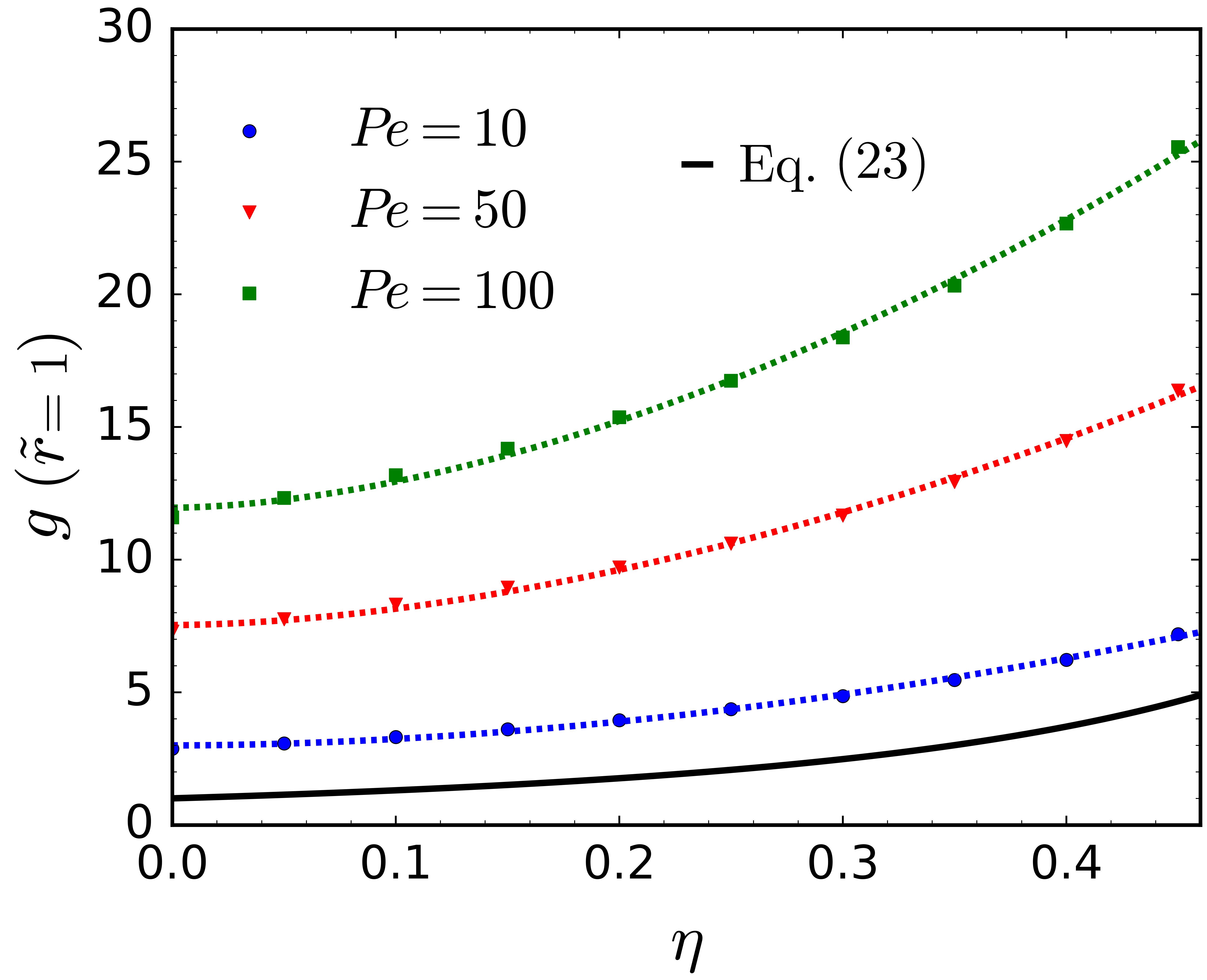}
	\caption{Value at contact of the pair correlation function as a function of the packing fraction $\eta,$ for several fixed values of the $\textrm{Pe}$ number. While points represent results of the introduced theoretical scheme, dashed lines indicate a fit to Eq. \eqref{fit_ETA} of the main text. At $\textrm{Pe}=10$ (blue) $\alpha=18.51,$ $\beta=1.87$ and $\gamma=2.99,$ at $\textrm{Pe}=50$ (red) $\alpha=45.39,$ $\beta=1.76$ and $\gamma=7.53,$ finally at $\textrm{Pe}=100$ (green) $\alpha=53.70,$ $\beta=1.73$ and $\gamma=11.95.$ The black full line indicates Eq. \eqref{CS_contact}, obtained from the Carnahan-Starling equation of state and holding in case $\textrm{Pe}=0.$ }
	\label{scaling_eta}
\end{figure}

In this Section we present predictions for the microscopic structure of a concentrated hard-sphere suspension under shear flow, obtained by using the framework of Section II.  We first compute the pair correlation function $g (\tilde{r})$ for several combinations of $\eta$ and $\textrm{Pe},$ and compare them to numerical data from previous numerical work by Morris and Katyal \citep{Morris_Katyal}. We then extract scaling laws for the value at contact of $g(\tilde{r})$ as a function of the $\textrm{Pe}$ number at fixed packing fraction $\eta,$ and  as a function of the packing fraction $\eta$ at fixed $\textrm{Pe}$ number. We finally investigate the effect of the shear flow on the structure factor.

\subsection{Comparison with numerical results from the literature}

By following the scheme introduced in Section II, we compute the $g(\tilde{r})$ function for several values of the packing fraction $\eta$ and of the $\textrm{Pe}$ number. In Fig. \ref{comparison_eta035} we use a red dashed line to plot our theoretically determined $g (\tilde{r})$ at fixed $\eta=0.30,$ in cases $\textrm{Pe}=25$ (a) and $\textrm{Pe}=1000$ (b), respectively. In the same figure we use points to present results form the simulations of Ref. \citep{Morris_Katyal}. These were obtained by using the Stokesian Dynamics technique in \citep{Morris_Katyal}. An excellent agreement between predictions of theory and results of numerical simulations can be observed for both values of the $\textrm{Pe}$ number. In particular, the theory (almost) correctly predicts the value of the pair correlation function at the contact distance from the reference particle $\tilde{r} =1.$ Moreover, the location and the value of a second (smaller) peak predicted by the theory are also in agreement with results from simulations. 

A graph similar to that of Fig. \ref{comparison_eta035} is presented in Fig. \ref{comparison_eta040} for the case $\eta=0.45.$ While a good qualitatively agreement between theoretical predictions and numerical findings is still found, a worse  quantitative agreement with respect to that of Fig. \ref{comparison_eta035} is observed. We attribute this slight disagreement to the growing importance that correlation functions involving more than two particles, e. g. three particle correlation functions, acquire upon increasing the packing fraction $\eta.$

Overall, Figs. \ref{comparison_eta035} and \ref{comparison_eta040} present a successful parameter-free test for the accuracy of our theoretical findings. We obtain correct results for the $g(\tilde{r})$ function in a range of the packing fraction which cannot be explored by (just) solving the pair Smoluchowski equation.

\subsection{Contact value $g (\tilde{r}=1)$ of the pair correlation function}

\begin{figure*}
	\centering
	\includegraphics[width = 0.9 \linewidth]{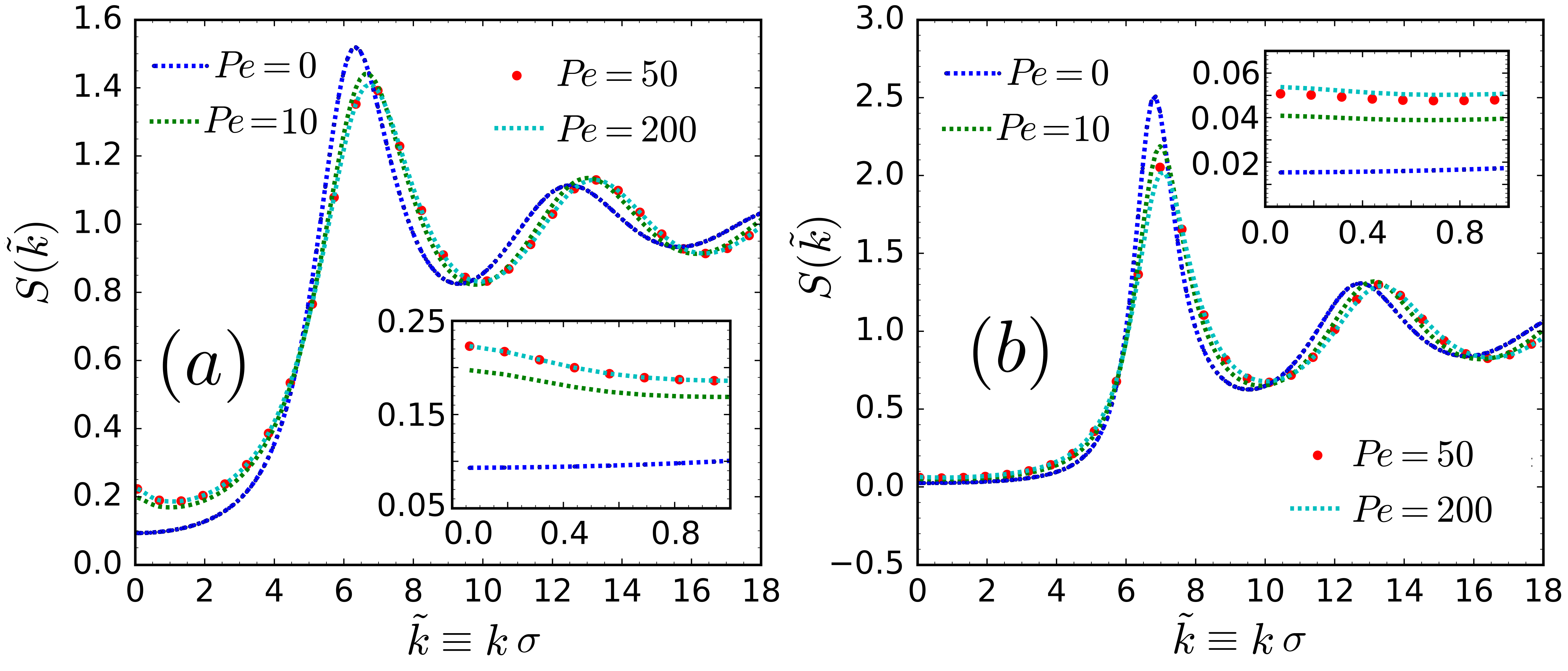}
	\caption{Structure factor $S (\tilde{k})$ of a hard-sphere colloidal suspension under shear flow, for several values of the $\textrm{Pe}$ number, at fixed packing fraction $\eta=0.30$ (a) and $\eta=0.45$ (b). In both cases, a consistent enhancement of $S(\tilde{k})$ at $\tilde{k} \to 0$ can be observed upon increasing the $\textrm{Pe}$ number. We argue (see main text) this behaviour to signal the onset of a shear-induced phase transition from the isotropic phase to a non-uniform one. The enhancement of the structure factor at small wavenumber is more pronounced at $\eta=0.30$ than at $\eta=0.45.$ At higher concentrations, indeed, the system presumably remains more uniform due to the higher density. Either in (a) and in (b) the inset shows a zoom in the region $\tilde{k} \in [0.0, 1.0].$}
	\label{structure_factor_Pe}
\end{figure*}

After our theoretical findings have been successfully compared with numerical results from the literature, we investigate how the shear flow affects the structural change experienced by a colloidal suspension when the packing fraction $\eta$ is increased. 

In Fig. \ref{structure_Pe}(a) we plot the $g(\tilde{r})$ function of a hard-sphere system in the absence of shear flow, i. e. at $\textrm{Pe}=0,$ for several increasing values of the packing fraction $\eta.$ As it is well-known, the shape of the $g(\tilde{r})$ function changes quite significantly by increasing $\eta.$ Indeed, while integral equation theories are typically not able to detect the onset of phase transitions, they can capture the variation occurring in the microscopic structure of a suspension when the density increases \citep{Brader_precursor}. We here investigate how such a structural variation is affected by the presence of a simple shear flow. To this aim we compute the $g (\tilde{r})$ function for several values of $\eta,$ at fixed $\textrm{Pe} \neq 0.$ In particular, in Fig.  \ref{structure_Pe}$(b)$ we plot $g(\tilde{r})$ for the same values of the packing fraction as in Fig. \ref{structure_Pe}(a), at fixed $\textrm{Pe}=50.$ We observe that, for all the considered values of $\eta,$ the value $g (\tilde{r}=1)$  of the pair correlation function at contact in case $\textrm{Pe} \neq 0$ is much larger than the same value obtained in case $\textrm{Pe} = 0.$ Moreover, the shear flow determines a shift of the radial position at which the second peak of the $g(\tilde{r})$ function is located. Indeed, while in Fig. \ref{structure_Pe}(a) the location of the second peak varies with $\eta,$ in  Fig. \ref{structure_Pe}(b) the second peak is always located at $\tilde{r}=2,$ for each value of the packing fraction $\eta.$

We will now extract scaling laws for $g (\tilde{r} =1)$ as a function of $\eta$ and $\textrm{Pe},$ respectively.

\subsubsection{Scaling of $g (\tilde{r}=1)$ with the P\'eclet number}

We here investigate how the value $g (\tilde{r} =1)$ of the pair correlation function at contact scales with the $\textrm{Pe}$ number, at fixed packing fraction $\eta.$ The first attempt to quantify this behaviour was performed by Brady and Morris \citep{brady_morris_1997}, who found for hard spheres the scaling relation $g (\tilde{r} = 1) \approx \textrm{Pe}.$  This result was successively revised by Morris and Katyal  \cite{Morris_Katyal}, who instead found $g(\tilde{r} =1 ) \approx \text{Pe}^{0.7}.$ While the scaling law of Brady and Morris was obtained by solving the pair Smoluchowski equation, the scaling law of Morris and Katyal was obtained through Stokesian dynamics simulations. 

In Fig. \ref{scaling_Pe}, we plot  $g (\tilde{r}=1)$ as a function of the $\textrm{Pe}$ number, at fixed $\eta = 0.30$ (a) and $\eta=0.45$ (b), respectively.  In both cases we find that our results can be fitted with the power law 
\begin{equation} \label{fit_Pe}
g(\tilde{r}=1) = \alpha \text{Pe}^{\beta},
\end{equation}
where the values of $\alpha$ and $\beta$ are specified in the caption of the figure. It is clear that our theory predicts a scaling law remarkably similar to the one obtained in the simulation study of Morris and Katyal \citep{Morris_Katyal}.
 
\subsubsection{Scaling of $g (\tilde{r}=1)$ with the packing fraction}

It is well-known that in the absence of shear flow, i. e. at $\textrm{Pe}=0,$ the value $g (\tilde{r} = 1)$ of the pair correlation function at contact provides the pressure $p$ of the uniform fluid as a function of its packing fraction $\eta \equiv \frac{4}{3} \pi (\sigma/2)^{3} \rho,$ through the relation \citep{Allen_Tildesley,Torquato_BOOK}
\begin{equation}
Z (\eta) = 1 + 4 \eta g (\tilde{r}=1),
\end{equation}
where $Z \equiv p / \rho k_{B} T$ is the so-called compressibility factor, $\rho$ is the number density, $T$ and $k_{B}$ are the (absolute) temperature and the Boltzmann constant, respectively. When the Carnahan-Starling relation $Z_{CS} (\eta) \equiv \big( 1 + \eta + \eta^{2} - \eta^{3} \big) / \big( 1 - \eta \big)^{3}$ is used to approximate the equation of state of the equilibrium system \citep{CS}, the functional dependence of $g (\tilde{r} =1)$ on the packing fraction $\eta$ is given by
\begin{equation} \label{CS_contact}
g (\tilde{r}=1) = \frac{1 - \eta/2}{(1- \eta)^{3}}.
\end{equation}
We here aim to investigate how the relation \eqref{CS_contact} is modified by the action of an external shear flow. We then investigate how $g (\tilde{r}=1)$  varies with $\eta,$ at fixed $\textrm{Pe}.$

In Fig. \ref{scaling_eta} we plot the value $g (\tilde{r}=1)$ of the pair correlation function at contact as a function of the packing fraction $\eta,$ for several fixed values of the $\textrm{Pe}$ number. In all cases, our theoretical results can be fitted to the following scaling law 
\begin{equation} \label{fit_ETA}
         g (\tilde{r}=1) = \alpha \eta^{\beta} + \gamma, 
\end{equation}
where $\alpha,$ $\beta$ and $\gamma$ are reported in the caption of the figure. 

To the best of our knowledge, a similar scaling law has never been reported in the literature. However, we believe such a relation could represent the first step towards a nonequilibrium equation of state for hard spheres under shear flow. For this reason, we hope our finding will inspire new studies to determine how the value at contact of the pair correlation function of a sheared colloidal suspension varies as a function of the packing fraction.

\subsection{Structure factor}

From the knowledge of the pair correlation function $g(\tilde{r}),$ the structure factor $S (k)$ of the system can be obtained through the relation \citep{Hansen_McDonald_BOOK}
\begin{equation} \label{structure_factor}
S (k) = 1 + \rho \int d \mathbf{r} g (r) e^{-i \mathbf{k} \cdot \mathbf{r}}.
\end{equation}
In other words, the structure factor $S (k)$ is given by the Fourier transform of the pair correlation function $g (\tilde{r}).$ 

Here we exploit Eq. \eqref{structure_factor} to study the effect of the shear flow on $S (k).$
In Fig. \ref{structure_factor_Pe} we plot the structure factor $S (k)$ for several values of the $\textrm{Pe}$ number, in case $\eta=0.30$ (a) and $\eta=0.45$ (b), respectively. We can observe the shear flow to cause several important effects. These include (i) a lowering of the main peak, (ii) an overall shift of the structure factor towards higher values of the wavevector, and (iii) an overall broadening of the first peak of the structure factor compared with the equilibrium conditions. These are all novel predictions that may stimulate experimental investigations in future work. 

Finally, Fig. \ref{structure_factor_Pe} shows that $S (k)$ consistently increases at $k\rightarrow 0,$ upon increasing the $\textrm{Pe}$ number. This effect is more visible at intermediate concentrations (see Fig. \ref{structure_factor_Pe}(a)) rather than at larger $\eta$ (see Fig. \ref{structure_factor_Pe}(b)).

A divergence of the structure factor at small wavenumber is known to occur in colloidal mixtures, in the absence of shear flow, and is associated with the physical instability of the mixture against phase separation \citep{Dzubiella_JCP_2002,mathias}. However, the same phenomenon has been observed also for one-component systems. Huang et al. \citep{Huang} reported an enhancement in the structure factor of water at small wavenumber under ambient conditions, and suggested this enhancement to signal the presence of anomalous density fluctuations. An explanation for the experimental observations of Huang et al. \citep{Huang} was successively provided by Overduin and Patey \citep{Overduin_Patey}. These authors showed that different local structural arrangements are present in water, which experience different effective interactions. The latter are attractive between molecules with similar local environments and repulsive between molecules with different local environments. The presence of attractive and repulsive interactions leads to concentration fluctuations which couple with density fluctuations and can account for the increase of the structure factor at low wavenumber.

We here invoke a similar mechanism to the one proposed by Overduin and Patey \citep{Overduin_Patey}, to explain the enhancement of $S (k)$ at $k \to 0$ predicted by our theory upon increasing the $\textrm{Pe}$ number. We argue the shear flow to induce structural heterogeneity in the colloidal suspension, which results in effective interactions and hence in density fluctuations. A phase-transition from the isotropic phase to a non-uniform one would then occur at a sufficiently large value of the $\textrm{Pe}$ number, as a result of the action of the external shear flow. Being associated with density fluctuations, the shear-induced transition could be of the kind described by Brazovskii \citep{Brazovski}. Finally, the transition is more likely to occur at intermediate concentrations than at larger concentrations where the system presumably remains more uniform due to the higher density.

\section{Conclusion}

In this paper, we introduced a theoretical framework to investigate the microscopic structure of concentrated hard-sphere colloidal suspensions subjected to a strong shear flow. We started by solving the pair Smoluchowski equation with shear, separately in the compressing and extensional sectors of the solid angle. To this aim, we followed a new analytical scheme based on intermediate asymptotics \citep{Banetta_Zaccone_PRE}. We then employed the obtained solution to construct a potential of mean force $u_{\textrm{eff}}$ containing the effect of the flow field on the pair correlation function, and inserted  $u_{\textrm{eff}}$  in the Percus-Yevick relation. We finally used the latter relation to solve the Ornstein-Zernike integral equation for a wide range of either the packing fraction $\eta$ and the P\'eclet number $\textrm{Pe}.$ Crucial to build the potential of mean force is to include hydrodynamic interactions, and treat them properly in the compressing and extensional sectors, respectively.

We obtained profiles for the pair correlation function which are in excellent agreement with  numerical results from Ref. \citep{Morris_Katyal} up to rather large values of $\eta,$ independently of the considered value of $\textrm{Pe}.$ We then extracted scaling laws for the value of the pair correlation function at contact as a function of the $\textrm{Pe}$ number at fixed $\eta,$ and as a function of the $\eta$ at fixed $\textrm{Pe}$ number. In the former case, we obtained a scaling law in agreement with the simulation study of Ref. \citep{Morris_Katyal}. In the latter case, we found a scaling law which could open the way for a non-equilibrium equation of state of strongly sheared liquids. Finally we employed our method to investigate the effect of the shear flow on the structure factor $S (k).$ 
The latter analysis reveals a consistent increase of $S (k)$ at $k \to 0,$ upon increasing the $\textrm{Pe}$ number. We argue this enhancement could signal the onset of a phase transition from the isotropic to a non-uniform state of the type discussed by Brazovskii \citep{Brazovski}, here induced by the external shear flow.

Several extensions of the work presented in this paper could be considered. While we have focused on the hard-sphere colloidal suspensions, the introduced theory holds for any (isotropic) interaction potential. It is then interesting to employ our scheme to investigate the effect of an external shear flow on the microscopic structure of suspensions of colloidal particles interacting through, e.g. Lennard-Jones or Yukawa (Debye-H\"uckel) potentials (the latter relevant for plasmas and electrolyte solutions). It is also interesting to include in the proposed framework correlation functions involving more than two particles, e. g. three-body correlation functions, in order to investigate the microscopic structure at larger values of the packing fraction, in the so-called dense regime. Finally, it is worthwhile to conduct further investigation to understand the nature of the shear-induced phase transition (possibly of the Brazovskii type) apparent at low-$k$ in the structure factor, predicted by our model. We aim to address these issues in future work.

\subsection*{Acknowledgments} 
C. A. gratefully acknowledges financial support from Syngenta AG. A.Z. acknowledges financial support from US Army Research Laboratory and US Army Research Office through contract nr. W911NF-19-2-0055. 

\appendix

\section{Angular averages}

We use the symbol $\left\langle \cdots \right\rangle_{i}$ to indicate angular averages. In particular, we use $i = ``c"$ to indicate the average over the compressing sectors of the solid angle, i.e. $\left\langle \cdots \right\rangle_{c} \equiv (2 \pi )^{-1} \int_{0}^{\pi} d \theta \sin \theta \big[ \int_{\pi/2}^{\pi} d \phi + \int_{3 \pi/2}^{2 \pi} d \phi \big],$ while we use $i = ``e"$ to indicate the average over the extensional sectors of the solid angle, i. e. $\left\langle \cdots \right\rangle_{e} \equiv (2 \pi )^{-1} \int_{0}^{\pi} d \theta \sin \theta \big[ \int_{0}^{\pi/2} d \phi + \int_{\pi}^{3 \pi/2} d \phi \big].$
The average of the pair correlation function $g (\mathbf{r})$ over the compressing sectors is hence defined as $g_{c} (r) \equiv \left\langle g (\mathbf{r}) \right\rangle_{c},$ while the average of $g (\mathbf{r})$ over the extensional sectors is defined as $g_{e} (r) \equiv \left\langle g (\mathbf{r}) \right\rangle_{e}.$ 

From the above it follows that 
\begin{equation}
\begin{aligned}
& \frac{g_{c} (r) + g_{e} (r)}{2} = \frac{\left\langle g (\mathbf{r}) \right\rangle_{c} + \left\langle g (\mathbf{r}) \right\rangle_{e}}{2} \\
&= \frac{1}{4 \pi} \int_{0}^{\pi} d \theta \sin \theta \bigg[ \int_{\pi/2}^{\pi} d \phi + \int_{3 \pi/2}^{2 \pi} d \phi \bigg] g (\mathbf{r}) \\
&+ \frac{1}{4 \pi}  \int_{0}^{\pi} d \theta \sin \theta \bigg[ \int_{0}^{\pi/2} d \phi + \int_{\pi}^{3 \pi/2} d \phi \bigg] g (\mathbf{r}) \\
&= \left\langle g (\mathbf{r}) \right\rangle,
\end{aligned}
\end{equation}
where we have defined $\left\langle \cdots \right\rangle \equiv (4 \pi )^{-1} \int_{0}^{\pi} d \theta \sin \theta \int_{0}^{2 \pi} d \phi.$ In Eq. \eqref{average} of Section II, we have introduced the function $g_{0} (r) \equiv \big( g_{c} (r) + g_{e} (r) \big) / 2$ which, as it is clear, corresponds to the average of the pair correlation function $g (\mathbf{r})$ over the full solid angle $\Omega \equiv (\theta, \phi)$ with $\theta \in [0, \pi]$ and $\phi \in [0, 2 \pi],$ respectively.

It is important to observe that, while the normalizing coefficient $(2 \pi)^{-1}$ is used in defining angular averages over the compressing and extensional sectors, the normalizing coefficient $(4 \pi)^{-1}$ is used in defining the angular average over the full solid angle.

\section{Solution of the pair Smoluchowski equation by intermediate asymptotics}

When averaging Eq. \eqref{Diluted_2_Body_Smoluchowski_Equation} we follow Refs. \citep{Banetta_Zaccone_PRE, Banetta_Yukawa} and assume the relative velocity $\mathbf{v}$ and the pair correlation function to be weakly correlated, such that
\begin{equation} 
\begin{aligned}
& \left\langle \tilde{\mathbf{v}} \cdot \tilde{\nabla} g (\tilde{\mathbf{r}}) \right\rangle_{i} \approx \left\langle \tilde{\mathbf{v}} \right\rangle_{i} \cdot \tilde{\nabla} g (\tilde{\mathbf{r}}),  \\
& \left\langle g (\tilde{\mathbf{r}}) \tilde{\nabla} \cdot \tilde{\mathbf{v}} \right\rangle_{i} \approx  g (\tilde{\mathbf{r}}) \left\langle \tilde{\nabla} \cdot \tilde{\mathbf{v}} \right\rangle_{i},
\end{aligned}
\end{equation}
where $i \in \big\{``c" , ``e" \big\}.$ The pair Smoluchowski equation \eqref{Diluted_2_Body_Smoluchowski_Equation} consequently becomes
\begin{equation} \label{final_formulation}
\begin{aligned}
& \epsilon \bigg[ G_{i} (\tilde{r}) \bigg( \frac{d^{2} g_{i} (\tilde{r})}{d \tilde{r}^{2}} + \frac{2}{\tilde{r}} \frac{d g_{i} (\tilde{r})}{d \tilde{r}}  \bigg) + \frac{d G_{i} (\tilde{r})}{d \tilde{r}} \frac{d g_{i} (\tilde{r})}{d \tilde{r}} \\ 
&+ g_{i} (\tilde{r}) \frac{d \tilde{u} (\tilde{r})}{d \tilde{r}} \frac{d G_{i} (\tilde{r})}{d \tilde{r}} 
+ G_{i} (\tilde{r}) \frac{d \tilde{u} (\tilde{r})}{d \tilde{r}} \frac{d g_{i} (\tilde{r})}{d \tilde{r}} \\
&+ G_{i} (\tilde{r}) \bigg( \frac{2}{\tilde{r}} \frac{d \tilde{u} (\tilde{r})}{d \tilde{r}} + \frac{d^{2} \tilde{u} (\tilde{r})}{d \tilde{r}^{2}} \bigg) g_{i} (\tilde{r}) \bigg] = 2 \left\langle \tilde{\mathbf{v}} \right\rangle_{i}  \frac{d g_{i} (\tilde{r})}{d \tilde{r}} \\
&+  2 g_{i} (\tilde{r}) \left\langle \tilde{\nabla} \cdot \tilde{\mathbf{v}} \right\rangle_{i} ,
\end{aligned}
\end{equation}
where $\epsilon \equiv 1 / \textrm{Pe},$ $i \in \big\{``c" , ``e" \big\}$ and $G_{c} (\tilde{r})$ and $G_{e} (\tilde{r})$ are given by Eq. \eqref{G_lubr_compressing} and Eq. \eqref{G_lubr_extensional}, respectively. In Eq. \eqref{final_formulation} we assume $\mathbf{v}$ to be given only by its radial component (see the first line of Eq. \eqref{velocity_HS} ), i. e. we assume $\mathbf{v} \approx v_{r}.$
From the definition of the angular averages given in Appendix A and (the first line of) Eq. \eqref{velocity_HS}, it follows that
\begin{equation}
\begin{aligned}
& \left\langle \tilde{\mathbf{v}} \right\rangle_{i} \approx \left\langle \tilde{v}_{r} \right\rangle_{i} = \alpha_{i} \big[ 1 - A (\tilde{r}) \big] \tilde{r}, \\
& \left\langle \tilde{\nabla} \cdot \tilde{\mathbf{v}} \right\rangle_{i} = \alpha_{i} \bigg[ 3 B (\tilde{r}) - 3 A (\tilde{r}) - \tilde{r} \frac{d A (\tilde{r})}{d \tilde{r}} \bigg],
\end{aligned}
\end{equation}
where $\alpha_{c} \equiv \left\langle \sin^{2} \theta \sin \phi \cos \phi \right\rangle_{c}  = - 2 /(3 \pi)$ and $\alpha_{e} \equiv  \left\langle \sin^{2} \theta \sin \phi \cos \phi \right\rangle_{e}  =  2 /(3 \pi).$  From $\alpha_{c}$ and $\alpha_{e}$ it is clear that the relative radial velocity between the particles is negative in the compressing sectors of the solid angle, while it is positive in the extensional sectors. It is important to notice that when $\mathbf{v}$ is averaged over the full solid angle, it is $\left\langle \mathbf{v} \right\rangle =  (4 \pi )^{-1} \int_{0}^{\pi} d \theta \sin \theta  \int_{0}^{2 \pi} d \phi \ v_{r} (\theta, \phi) =0.$

In order to fully specify the problem, Eq \eqref{final_formulation} has to be supplemented with two boundary conditions. The first of these is a no-flux condition at $\tilde{r}=\tilde{r}_{c},$ 
\begin{equation} \label{BC_flux}
\bigg[ G_{i} (\tilde{r}) \frac{d g ( \tilde{r})}{d \tilde{r}} + \bigg( G_{i} (\tilde{r}) \frac{d \tilde{u} (\tilde{r})}{d \tilde{r}} - 2 \textrm{Pe} \left\langle \tilde{\mathbf{v}} \right\rangle_{i} \bigg) g_{i} (\tilde{r} ) \bigg] \Bigr|_{\substack{\tilde{r} = \tilde{r}_{c}}}=0,
\end{equation}
where $i \in \big\{``c" , ``e" \big\}$ and $\tilde{r}_{c}$ is a value of radial distance sufficiently close to the reference particle. We here take $\tilde{r}_{c} = 1 + 5 \times 10^{5}.$ The second boundary condition is instead
\begin{equation}
g_{i} (\tilde{r} \to \infty) =1.
\end{equation}

Eq. \eqref{final_formulation} is an example of so-called \textit{singular perturbation problem}, i. e. an ordinary differential equation with  perturbation parameter $\epsilon$ linked to the highest order derivative. In this case the problem can be approached by using the \textit{boundary layer} theory \citep{Bender_Orszag}. The approach consists of the evaluation of two different series in two different regions of the domain: the \textit{outer layer} where the solution is slowly varying with $\tilde{r},$ and the \textit{inner layer}, also known as the \textit{boundary layer}, where the solution is rapidly varying with $\tilde{r}.$

In the outer layer we write
\begin{equation} \label{outer_expansion}
g_{i}^{\textrm{out}} (\tilde{r}) \approx g_{0,i}^{\textrm{out}} (\tilde{r}) + \epsilon g_{1,i}^{\textrm{out}} (\tilde{r}) + \mathcal{O} (\epsilon^{2}),
\end{equation}
where $\epsilon \equiv 1 / \textrm{Pe}.$ To introduce the power series in the inner layer a change of variable, called the \textit{inner transformation}, in Eq. \eqref{final_formulation} needs to be considered \citep{Banetta_Zaccone_PRE}. In our case, the inner transformation reads
\begin{equation}
\xi \equiv \frac{\tilde{r} - \tilde{r}_{c}}{\delta (\epsilon)},
\end{equation}
where $\delta (\epsilon)$ is the order of magnitude of the width of the inner layer. Using the method of \textit{dominant balancing}, in Ref. \citep{Banetta_Zaccone_PRE} it was shown that $\delta (\epsilon) \approx \epsilon.$ The power expansion in the inner layer can then be written as
\begin{equation} \label{inner_expansion}
g_{i}^{\textrm{in}} (\xi) \approx g_{0,i}^{\textrm{in}} (\xi) + \epsilon g_{1,i}^{\textrm{in}} (\xi) + \mathcal{O} (\epsilon^{2}),
\end{equation}
where $\epsilon \equiv 1 / \textrm{Pe}.$

As showed in Refs. \citep{Banetta_Zaccone_PRE,Banetta_Yukawa}, the $g_{0,i}^{\textrm{out}} (\tilde{r})$ and $g_{1,i}^{\textrm{out}} (\tilde{r})$ appearing in the expansion Eq. \eqref{outer_expansion} are given by
\begin{equation}
\begin{aligned}
& g_{0,i}^{\textrm{out}} (\tilde{r}) = \frac{1}{1 - A (\tilde{r})} \exp \bigg[ \int_{\tilde{r}}^{\infty} d \tilde{r}^{'}  \ \frac{3B (\tilde{r}^{'} ) - 3A (\tilde{r}^{'} ) }{\tilde{r}^{'}  -  \tilde{r}^{'}  A (\tilde{r}^{'} )}   \bigg],
\end{aligned}
\end{equation}
and
\begin{equation} \label{Eq8888}
\begin{aligned}
& g_{1,i}^{\textrm{out}} (\tilde{r}) = - g_{0,i}^{\textrm{out}} (\tilde{r}) \int_{\tilde{r}}^{\infty} \frac{d \tilde{r}^{'} }{2 \left\langle \tilde{\mathbf{v}} \right\rangle_{i}} \bigg\{ G_{i} ( \tilde{r}^{'} ) \bigg[ \big( Y ( \tilde{r}^{'}) \big)^{2} + \frac{d Y ( \tilde{r}^{'} )}{d \tilde{r}^{'}} \\
&+ \bigg( \frac{2}{\tilde{r}^{'}} + \frac{d \tilde{u} (\tilde{r}^{'})}{d \tilde{r}} \bigg) Y (\tilde{r}^{'}) + \frac{d^{2} \tilde{u} (\tilde{r}^{'})}{d \tilde{r}^{2}} + \frac{2}{\tilde{r}^{'}} \frac{d \tilde{u} (\tilde{r}^{'}) }{d \tilde{r}^{'}} \bigg] \\
& + \frac{d G_{i} (\tilde{r}^{'})}{d \tilde{r}^{'}} \bigg( Y (\tilde{r}^{'}) + \frac{d \tilde{u} (\tilde{r}^{'})}{d \tilde{r}^{'}} \bigg) \bigg\},
\end{aligned}
\end{equation}
respectively. In \eqref{Eq8888} we have defined $Y (\tilde{r}) \equiv - \left\langle \tilde{\nabla} \cdot \tilde{\mathbf{v}} \right\rangle_{i} / \left\langle \tilde{\mathbf{v}} \right\rangle_{i}.$

As showed in Refs.  \citep{Banetta_Zaccone_PRE,Banetta_Yukawa}, the $g_{0,i}^{\textrm{in}} (\xi)$ and $g_{1,i}^{\textrm{in}} (\xi)$ appearing in the expansion Eq. \eqref{inner_expansion} are given by
\begin{equation} \label{eq99998}
\begin{aligned}
& g_{0,i}^{\textrm{in}} (\xi) = C_{1} + C_{0} \int_{0}^{\xi} d \xi^{'} \exp \bigg[ \int_{0}^{\xi^{'}} 2 \frac{\left\langle \tilde{\mathbf{v}} (\epsilon=0) \right\rangle_{i}}{ G (\epsilon=0)} d \xi \bigg], 
\end{aligned}
\end{equation}
and
\begin{equation} \label{eq9999}
\begin{aligned}
&  g_{1,i}^{\textrm{in}} (\xi) = C_{3} + \int_{0}^{\xi}  d \xi^{'} \bigg\{ C_{2} - \int_{0}^{\xi^{'}}  d \xi^{''} \bigg[ \bigg( \frac{2}{\xi^{''} \epsilon + \tilde{r}_{c}} + W(\xi^{''}) \\
& + \frac{G_{r,i} (\xi^{''})}{G (\xi^{''})} \bigg) \frac{d g_{0,i}^{\textrm{in}} (\xi^{''})}{d \xi^{''}} - 2 \frac{\left\langle \tilde{\nabla}_{\xi^{''}} \cdot \tilde{\mathbf{v}} (\xi^{''}) \right\rangle_{i}}{G (\xi^{''})} g_{0,i}^{\textrm{in}} (\xi^{''}) \bigg] \times \\
& \times \exp \bigg( -2 \int_{0}^{\xi^{'}}  d \xi \frac{\left\langle \tilde{\mathbf{v}} (\xi) \right\rangle_{i}}{G (\xi)} \bigg) \bigg\} \exp \bigg( - 2 \int_{0}^{\xi^{'}}  d \xi \frac{\left\langle \tilde{\mathbf{v}} (\xi) \right\rangle_{i}}{G (\xi)} \bigg) ,
\end{aligned}
\end{equation}
respectively. 
In Eq. \eqref{eq9999} we have defined $W (\xi) \equiv (d \tilde{u} (\xi)/ d \xi ) / \delta$ and $G_{r,i} (\xi) = \delta^{-1} (d G_{i} (\xi) / d \xi).$
 
The final step to obtain the analytical solution of Eq. \eqref{final_formulation} is the evaluation of the integration constants $C_{0}, \ C_{1}, \ C_{2}$ and $C_{3}$ present in Eq. \eqref{eq99998} and Eq. \eqref{eq9999}.

Since our problem contains four (unknown) integration constants, four conditions are needed to determine them. The first of these conditions is  the condition of zero flux at the reference particle surface Eq. \eqref{BC_flux}. The other three are
\begin{equation}
    \begin{aligned}
&     g_{i}^{\textrm{out}} (\tilde{r} = \tilde{r}_{c} + \epsilon) = g_{i}^{\textrm{in}} (\tilde{r} = \tilde{r}_{c} + \epsilon) ,\\
 &    \frac{d g_{i}^{\textrm{out}} (\tilde{r}) }{ d \tilde{r} }  \Bigr|_{\substack{\tilde{r} = \tilde{r}_{c} + \epsilon}} = \frac{d g_{i}^{\textrm{in}} (\tilde{r}) }{ d \tilde{r} }  \Bigr|_{\substack{\tilde{r} = \tilde{r}_{c} + \epsilon}}, \\
 & \frac{d^{2} g_{i}^{\textrm{out}} (\tilde{r}) }{ d \tilde{r}^{2} }  \Bigr|_{\substack{\tilde{r} = \tilde{r}_{c} + \epsilon}} = \frac{d^{2} g_{i}^{\textrm{in}} (\tilde{r}) }{ d \tilde{r}^{2} }  \Bigr|_{\substack{\tilde{r} = \tilde{r}_{c} + \epsilon}}.
    \end{aligned}
\end{equation}
These are derived by the so-called \textit{patching} procedure \citep{Bender_Orszag}, and specify that the inner solution must match the outer solution at the boundary layer $\tilde{r} = \tilde{r}_{c} + \epsilon,$ in a smooth (differentiable as many times as possible) way.

\bibliography{references}

\begin{thebibliography}{34}
\expandafter\ifx\csname natexlab\endcsname\relax\def\natexlab#1{#1}\fi
\expandafter\ifx\csname bibnamefont\endcsname\relax
  \def\bibnamefont#1{#1}\fi
\expandafter\ifx\csname bibfnamefont\endcsname\relax
  \def\bibfnamefont#1{#1}\fi
\expandafter\ifx\csname citenamefont\endcsname\relax
  \def\citenamefont#1{#1}\fi
\expandafter\ifx\csname url\endcsname\relax
  \def\url#1{\texttt{#1}}\fi
\expandafter\ifx\csname urlprefix\endcsname\relax\def\urlprefix{URL }\fi
\providecommand{\bibinfo}[2]{#2}
\providecommand{\eprint}[2][]{\url{#2}}

\bibitem[{\citenamefont{Hansen and McDonald}(2006)}]{Hansen_McDonald_BOOK}
\bibinfo{author}{\bibfnamefont{J.}~\bibnamefont{Hansen}} \bibnamefont{and}
  \bibinfo{author}{\bibfnamefont{I.}~\bibnamefont{McDonald}},
  \emph{\bibinfo{title}{Theory of Simple Liquids}}
  (\bibinfo{publisher}{Elsevier Science, New York}, \bibinfo{year}{2006}).

\bibitem[{\citenamefont{Allen and Tildesley}(2017)}]{Allen_Tildesley}
\bibinfo{author}{\bibfnamefont{M.~P.} \bibnamefont{Allen}} \bibnamefont{and}
  \bibinfo{author}{\bibfnamefont{D.~J.} \bibnamefont{Tildesley}},
  \emph{\bibinfo{title}{Computer Simulation of Liquids}}
  (\bibinfo{publisher}{Oxford University Press}, \bibinfo{year}{2017}).

\bibitem[{\citenamefont{Caccamo}(1996)}]{caccamoIET}
\bibinfo{author}{\bibfnamefont{C.}~\bibnamefont{Caccamo}},
  \bibinfo{journal}{Physics Reports} \textbf{\bibinfo{volume}{274}},
  \bibinfo{pages}{1} (\bibinfo{year}{1996}).

\bibitem[{\citenamefont{Brader et~al.}(2008)\citenamefont{Brader, Cates, and
  Fuchs}}]{PhysRevLett.101.138301}
\bibinfo{author}{\bibfnamefont{J.~M.} \bibnamefont{Brader}},
  \bibinfo{author}{\bibfnamefont{M.~E.} \bibnamefont{Cates}}, \bibnamefont{and}
  \bibinfo{author}{\bibfnamefont{M.}~\bibnamefont{Fuchs}},
  \bibinfo{journal}{Phys. Rev. Lett.} \textbf{\bibinfo{volume}{101}},
  \bibinfo{pages}{138301} (\bibinfo{year}{2008}).

\bibitem[{\citenamefont{Fuchs and Cates}(2002)}]{PhysRevLett.89.248304}
\bibinfo{author}{\bibfnamefont{M.}~\bibnamefont{Fuchs}} \bibnamefont{and}
  \bibinfo{author}{\bibfnamefont{M.~E.} \bibnamefont{Cates}},
  \bibinfo{journal}{Phys. Rev. Lett.} \textbf{\bibinfo{volume}{89}},
  \bibinfo{pages}{248304} (\bibinfo{year}{2002}).

\bibitem[{\citenamefont{Preziosi et~al.}(2017)\citenamefont{Preziosi, Perazzo,
  Tomaiuolo, Pipich, Danino, Paduano, and Guido}}]{Preziosi}
\bibinfo{author}{\bibfnamefont{V.}~\bibnamefont{Preziosi}},
  \bibinfo{author}{\bibfnamefont{A.}~\bibnamefont{Perazzo}},
  \bibinfo{author}{\bibfnamefont{G.}~\bibnamefont{Tomaiuolo}},
  \bibinfo{author}{\bibfnamefont{V.}~\bibnamefont{Pipich}},
  \bibinfo{author}{\bibfnamefont{D.}~\bibnamefont{Danino}},
  \bibinfo{author}{\bibfnamefont{L.}~\bibnamefont{Paduano}}, \bibnamefont{and}
  \bibinfo{author}{\bibfnamefont{S.}~\bibnamefont{Guido}},
  \bibinfo{journal}{Soft Matter} \textbf{\bibinfo{volume}{13}},
  \bibinfo{pages}{5696} (\bibinfo{year}{2017}).

\bibitem[{\citenamefont{Wu et~al.}(2010)\citenamefont{Wu, Tsoutsoura, Lattuada,
  Zaccone, and Morbidelli}}]{doi:10.1021/la902800x}
\bibinfo{author}{\bibfnamefont{H.}~\bibnamefont{Wu}},
  \bibinfo{author}{\bibfnamefont{A.}~\bibnamefont{Tsoutsoura}},
  \bibinfo{author}{\bibfnamefont{M.}~\bibnamefont{Lattuada}},
  \bibinfo{author}{\bibfnamefont{A.}~\bibnamefont{Zaccone}}, \bibnamefont{and}
  \bibinfo{author}{\bibfnamefont{M.}~\bibnamefont{Morbidelli}},
  \bibinfo{journal}{Langmuir} \textbf{\bibinfo{volume}{26}},
  \bibinfo{pages}{2761} (\bibinfo{year}{2010}), \bibinfo{note}{pMID: 19845347}.

\bibitem[{\citenamefont{Vermant and Solomon}(2005)}]{Vermant_2005}
\bibinfo{author}{\bibfnamefont{J.}~\bibnamefont{Vermant}} \bibnamefont{and}
  \bibinfo{author}{\bibfnamefont{M.~J.} \bibnamefont{Solomon}},
  \bibinfo{journal}{Journal of Physics: Condensed Matter}
  \textbf{\bibinfo{volume}{17}}, \bibinfo{pages}{R187} (\bibinfo{year}{2005}).

\bibitem[{\citenamefont{Dhont}(1996)}]{Dhont_BOOK}
\bibinfo{author}{\bibfnamefont{J.~K.~G.} \bibnamefont{Dhont}},
  \emph{\bibinfo{title}{An introduction to the dynamics of colloids}}
  (\bibinfo{publisher}{Elsevier, Amsterdam}, \bibinfo{year}{1996}).

\bibitem[{\citenamefont{Batchelor and Green}(1972)}]{batchelor_green_1972}
\bibinfo{author}{\bibfnamefont{G.~K.} \bibnamefont{Batchelor}}
  \bibnamefont{and} \bibinfo{author}{\bibfnamefont{J.~T.} \bibnamefont{Green}},
  \bibinfo{journal}{Journal of Fluid Mechanics} \textbf{\bibinfo{volume}{56}},
  \bibinfo{pages}{401–427} (\bibinfo{year}{1972}).

\bibitem[{\citenamefont{Brady and Morris}(1997)}]{brady_morris_1997}
\bibinfo{author}{\bibfnamefont{J.~F.} \bibnamefont{Brady}} \bibnamefont{and}
  \bibinfo{author}{\bibfnamefont{J.~F.} \bibnamefont{Morris}},
  \bibinfo{journal}{Journal of Fluid Mechanics} \textbf{\bibinfo{volume}{348}},
  \bibinfo{pages}{103–139} (\bibinfo{year}{1997}).

\bibitem[{\citenamefont{Dhont}(1989)}]{dhont_1989}
\bibinfo{author}{\bibfnamefont{J.~K.~G.} \bibnamefont{Dhont}},
  \bibinfo{journal}{Journal of Fluid Mechanics} \textbf{\bibinfo{volume}{204}},
  \bibinfo{pages}{421–431} (\bibinfo{year}{1989}).

\bibitem[{\citenamefont{B\l{}awzdziewicz and Szamel}(1993)}]{Szamel}
\bibinfo{author}{\bibfnamefont{J.}~\bibnamefont{B\l{}awzdziewicz}}
  \bibnamefont{and} \bibinfo{author}{\bibfnamefont{G.}~\bibnamefont{Szamel}},
  \bibinfo{journal}{Phys. Rev. E} \textbf{\bibinfo{volume}{48}},
  \bibinfo{pages}{4632} (\bibinfo{year}{1993}).

\bibitem[{\citenamefont{Schwarzl and Hess}(1986)}]{Schwarzl}
\bibinfo{author}{\bibfnamefont{J.~F.} \bibnamefont{Schwarzl}} \bibnamefont{and}
  \bibinfo{author}{\bibfnamefont{S.}~\bibnamefont{Hess}},
  \bibinfo{journal}{Phys. Rev. A} \textbf{\bibinfo{volume}{33}},
  \bibinfo{pages}{4277} (\bibinfo{year}{1986}).

\bibitem[{\citenamefont{Ronis}(1984)}]{Ronis}
\bibinfo{author}{\bibfnamefont{D.}~\bibnamefont{Ronis}},
  \bibinfo{journal}{Phys. Rev. A} \textbf{\bibinfo{volume}{29}},
  \bibinfo{pages}{1453} (\bibinfo{year}{1984}),
  \urlprefix\url{https://link.aps.org/doi/10.1103/PhysRevA.29.1453}.

\bibitem[{\citenamefont{de~Kruif et~al.}(1990)\citenamefont{de~Kruif, van~der
  Werff, Johnson, and May}}]{Kruif}
\bibinfo{author}{\bibfnamefont{C.~G.} \bibnamefont{de~Kruif}},
  \bibinfo{author}{\bibfnamefont{J.~C.} \bibnamefont{van~der Werff}},
  \bibinfo{author}{\bibfnamefont{S.~J.} \bibnamefont{Johnson}},
  \bibnamefont{and} \bibinfo{author}{\bibfnamefont{R.~P.} \bibnamefont{May}},
  \bibinfo{journal}{Physics of Fluids A: Fluid Dynamics}
  \textbf{\bibinfo{volume}{2}}, \bibinfo{pages}{1545} (\bibinfo{year}{1990}).

\bibitem[{\citenamefont{Clark and Ackerson}(1980)}]{Clark}
\bibinfo{author}{\bibfnamefont{N.~A.} \bibnamefont{Clark}} \bibnamefont{and}
  \bibinfo{author}{\bibfnamefont{B.~J.} \bibnamefont{Ackerson}},
  \bibinfo{journal}{Phys. Rev. Lett.} \textbf{\bibinfo{volume}{44}},
  \bibinfo{pages}{1005} (\bibinfo{year}{1980}).

\bibitem[{\citenamefont{Ackerson}(1990)}]{Ackerson}
\bibinfo{author}{\bibfnamefont{B.~J.} \bibnamefont{Ackerson}},
  \bibinfo{journal}{Journal of Rheology} \textbf{\bibinfo{volume}{34}},
  \bibinfo{pages}{553} (\bibinfo{year}{1990}).

\bibitem[{\citenamefont{Banetta and Zaccone}(2019)}]{Banetta_Zaccone_PRE}
\bibinfo{author}{\bibfnamefont{L.}~\bibnamefont{Banetta}} \bibnamefont{and}
  \bibinfo{author}{\bibfnamefont{A.}~\bibnamefont{Zaccone}},
  \bibinfo{journal}{Phys. Rev. E} \textbf{\bibinfo{volume}{99}},
  \bibinfo{pages}{052606} (\bibinfo{year}{2019}).

\bibitem[{\citenamefont{Banetta and Zaccone}(2020)}]{Banetta_Yukawa}
\bibinfo{author}{\bibfnamefont{L.}~\bibnamefont{Banetta}} \bibnamefont{and}
  \bibinfo{author}{\bibfnamefont{A.}~\bibnamefont{Zaccone}},
  \bibinfo{journal}{Colloid and Polymer Science}
  \textbf{\bibinfo{volume}{298}}, \bibinfo{pages}{761} (\bibinfo{year}{2020}).

\bibitem[{\citenamefont{Morris and Katyal}(2002)}]{Morris_Katyal}
\bibinfo{author}{\bibfnamefont{J.~F.} \bibnamefont{Morris}} \bibnamefont{and}
  \bibinfo{author}{\bibfnamefont{B.}~\bibnamefont{Katyal}},
  \bibinfo{journal}{Physics of Fluids} \textbf{\bibinfo{volume}{14}},
  \bibinfo{pages}{1920} (\bibinfo{year}{2002}).

\bibitem[{\citenamefont{{Brazovski{\v{i}}}}(1975)}]{Brazovski}
\bibinfo{author}{\bibfnamefont{S.~A.} \bibnamefont{{Brazovski{\v{i}}}}},
  \bibinfo{journal}{Soviet Journal of Experimental and Theoretical Physics}
  \textbf{\bibinfo{volume}{41}}, \bibinfo{pages}{85} (\bibinfo{year}{1975}).

\bibitem[{\citenamefont{Brader}(2010)}]{Brader_2010}
\bibinfo{author}{\bibfnamefont{J.~M.} \bibnamefont{Brader}},
  \bibinfo{journal}{Journal of Physics: Condensed Matter}
  \textbf{\bibinfo{volume}{22}}, \bibinfo{pages}{363101}
  (\bibinfo{year}{2010}).

\bibitem[{\citenamefont{Bender and Orszag}(1999)}]{Bender_Orszag}
\bibinfo{author}{\bibfnamefont{C.~M.} \bibnamefont{Bender}} \bibnamefont{and}
  \bibinfo{author}{\bibfnamefont{S.~A.} \bibnamefont{Orszag}},
  \emph{\bibinfo{title}{Advanced mathematical methods for scientists and
  \\engineers I: Asymptotic methods and perturbation theory}}
  (\bibinfo{publisher}{Springer Science and Business Media, New York},
  \bibinfo{year}{1999}).

\bibitem[{\citenamefont{Mulero}(2008)}]{Mulero}
\bibinfo{author}{\bibfnamefont{A.}~\bibnamefont{Mulero}},
  \emph{\bibinfo{title}{Theory and Simulation of Hard-Sphere\\ Fluids and
  Related Systems}} (\bibinfo{publisher}{volume 753 of Lecture Notes in
  Physics, Berlin Springer Verlag}, \bibinfo{year}{2008}).

\bibitem[{\citenamefont{Adler}(1981)}]{adler}
\bibinfo{author}{\bibfnamefont{P.}~\bibnamefont{Adler}},
  \bibinfo{journal}{Journal of Colloid and Interface Science}
  \textbf{\bibinfo{volume}{84}}, \bibinfo{pages}{461} (\bibinfo{year}{1981}).

\bibitem[{\citenamefont{Melis et~al.}(1999)\citenamefont{Melis, Verduyn,
  Storti, Morbidelli, and Bałdyga}}]{Melis}
\bibinfo{author}{\bibfnamefont{S.}~\bibnamefont{Melis}},
  \bibinfo{author}{\bibfnamefont{M.}~\bibnamefont{Verduyn}},
  \bibinfo{author}{\bibfnamefont{G.}~\bibnamefont{Storti}},
  \bibinfo{author}{\bibfnamefont{M.}~\bibnamefont{Morbidelli}},
  \bibnamefont{and} \bibinfo{author}{\bibfnamefont{J.}~\bibnamefont{Bałdyga}},
  \bibinfo{journal}{AIChE Journal} \textbf{\bibinfo{volume}{45}},
  \bibinfo{pages}{1383} (\bibinfo{year}{1999}).

\bibitem[{\citenamefont{Brader}(2008)}]{Brader_precursor}
\bibinfo{author}{\bibfnamefont{J.~M.} \bibnamefont{Brader}},
  \bibinfo{journal}{The Journal of Chemical Physics}
  \textbf{\bibinfo{volume}{128}}, \bibinfo{pages}{104503}
  (\bibinfo{year}{2008}).

\bibitem[{\citenamefont{Torquato}(2002)}]{Torquato_BOOK}
\bibinfo{author}{\bibfnamefont{S.}~\bibnamefont{Torquato}},
  \emph{\bibinfo{title}{Random Heterogeneous Materials:\\ Microstructure and
  Macroscopic Properties}} (\bibinfo{publisher}{Springer-Verlag, New York},
  \bibinfo{year}{2002}).

\bibitem[{\citenamefont{Carnahan and Starling}(1969)}]{CS}
\bibinfo{author}{\bibfnamefont{N.~F.} \bibnamefont{Carnahan}} \bibnamefont{and}
  \bibinfo{author}{\bibfnamefont{K.~E.} \bibnamefont{Starling}},
  \bibinfo{journal}{The Journal of Chemical Physics}
  \textbf{\bibinfo{volume}{51}}, \bibinfo{pages}{635} (\bibinfo{year}{1969}).

\bibitem[{\citenamefont{Dzubiella et~al.}(2002)\citenamefont{Dzubiella, Likos,
  and Löwen}}]{Dzubiella_JCP_2002}
\bibinfo{author}{\bibfnamefont{J.}~\bibnamefont{Dzubiella}},
  \bibinfo{author}{\bibfnamefont{C.~N.} \bibnamefont{Likos}}, \bibnamefont{and}
  \bibinfo{author}{\bibfnamefont{H.}~\bibnamefont{Löwen}},
  \bibinfo{journal}{The Journal of Chemical Physics}
  \textbf{\bibinfo{volume}{116}}, \bibinfo{pages}{9518} (\bibinfo{year}{2002}).

\bibitem[{\citenamefont{Schmidt}(2001)}]{mathias}
\bibinfo{author}{\bibfnamefont{M.}~\bibnamefont{Schmidt}},
  \bibinfo{journal}{Phys. Rev. E} \textbf{\bibinfo{volume}{63}},
  \bibinfo{pages}{050201} (\bibinfo{year}{2001}),
  \urlprefix\url{https://link.aps.org/doi/10.1103/PhysRevE.63.050201}.

\bibitem[{\citenamefont{Huang et~al.}(2009)\citenamefont{Huang, Wikfeldt,
  Tokushima, Nordlund, Harada, Bergmann, Niebuhr, Weiss, Horikawa, Leetmaa
  et~al.}}]{Huang}
\bibinfo{author}{\bibfnamefont{C.}~\bibnamefont{Huang}},
  \bibinfo{author}{\bibfnamefont{K.~T.} \bibnamefont{Wikfeldt}},
  \bibinfo{author}{\bibfnamefont{T.}~\bibnamefont{Tokushima}},
  \bibinfo{author}{\bibfnamefont{D.}~\bibnamefont{Nordlund}},
  \bibinfo{author}{\bibfnamefont{Y.}~\bibnamefont{Harada}},
  \bibinfo{author}{\bibfnamefont{U.}~\bibnamefont{Bergmann}},
  \bibinfo{author}{\bibfnamefont{M.}~\bibnamefont{Niebuhr}},
  \bibinfo{author}{\bibfnamefont{T.~M.} \bibnamefont{Weiss}},
  \bibinfo{author}{\bibfnamefont{Y.}~\bibnamefont{Horikawa}},
  \bibinfo{author}{\bibfnamefont{M.}~\bibnamefont{Leetmaa}},
  \bibnamefont{et~al.}, \bibinfo{journal}{Proceedings of the National Academy
  of Sciences} \textbf{\bibinfo{volume}{106}}, \bibinfo{pages}{15214}
  (\bibinfo{year}{2009}).

\bibitem[{\citenamefont{Overduin and Patey}(2012)}]{Overduin_Patey}
\bibinfo{author}{\bibfnamefont{S.~D.} \bibnamefont{Overduin}} \bibnamefont{and}
  \bibinfo{author}{\bibfnamefont{G.~N.} \bibnamefont{Patey}},
  \bibinfo{journal}{The Journal of Physical Chemistry B}
  \textbf{\bibinfo{volume}{116}}, \bibinfo{pages}{12014}
  (\bibinfo{year}{2012}).

\end{thebibliography}

\end{document}